\documentclass[twocolumn]{autart}    
\usepackage{graphicx}          
\usepackage[dvips]{epsfig}    
                               
\usepackage{amsmath,amssymb,amsfonts}
\usepackage{mathrsfs}
\usepackage[algo2e,ruled,vlined,linesnumbered,norelsize]{algorithm2e}
\usepackage{xcolor}
\usepackage{multirow}
\usepackage{makecell}
\usepackage{paralist}
\usepackage{float}

\usepackage{epstopdf}
\usepackage[utf8]{inputenc}
\usepackage{comment}
\usepackage{graphicx}
\usepackage{tikz}
\newcommand\mathsc[1]{\mbox{\textsc{\footnotesize #1}}}
\usetikzlibrary{shapes, positioning}
\usetikzlibrary{trees, calc, shapes.geometric, positioning} 

\DeclareMathOperator*{\BigAnd}{\bigwedge}
\DeclareMathOperator*{\BigOr}{\bigvee}

\newcommand\Real{\mathbb{R}}
\newcommand\Nat{\mathbb{N}}

\newcommand\setof[1]{\left\{ #1 \right\}}
\newcommand\conv{\mathsf{co}}
\newcommand\grad \nabla 
\newcommand\commentOut[1]{}

\begin{document}

\begin{frontmatter}

\title{Synthesizing Min-Max Control Barrier Functions For Switched Affine Systems \thanksref{footnoteinfo}} 

\thanks[footnoteinfo]{Will be added in subsequent version.}

\author[Boulder]{Sara Kamali}\ead{sara.kamali@colorado.edu},               
\author[Louvain]{Guillaume O. Berger}\ead{gberger@uclouvain.be},  
\author[Boulder]{Sriram Sankaranarayanan}\ead{srirams@colorado.edu}

\address[Boulder]{University of Colorado Boulder, USA.}             
\address[Louvain]{UC Louvain, Belgium.}        


\begin{abstract}  
We study the problem of synthesizing non-smooth control barrier functions (CBFs) for continuous-time switched affine systems. Switched affine systems are defined by a set of affine dynamical modes, wherein the control consists of a state-based switching signal that determines the current operating mode. The control barrier functions seek to maintain the system state inside a control invariant set that excludes a given set of unsafe states.  We consider CBFs that take the form of pointwise minima and maxima over a finite set of affine functions. Our approach uses ideas from nonsmooth analysis to formulate conditions for min- and max- affine control barrier functions. We show how a feedback switching law can be extracted from a given CBF. Next, we show how to automate the process of synthesizing CBFs given a system description through a tree-search algorithm inspired by branch-and-cut methods from combinatorial optimization. Finally, we demonstrate our approach on a series of interesting examples of switched affine systems. 
\end{abstract}

\end{frontmatter}

\section{Introduction}\label{sec:Introduction}

We study safety enforcement for switched affine systems through piecewise affine (min-max) control barrier functions. 
Designing safe controllers is a problem of critical importance for applications such as  autonomous vehicles, robotics, and healthcare applications. Safety ensures that system trajectories remain within a predefined safe region, preventing undesired or hazardous outcomes.  Control Barrier Functions (CBFs) \cite{ames2019control} form a well-established approach to safety enforcement for autonomous systems. Informally, CBFs are functions over the state-space such that the set of states wherein the CBF is non-positive are control invariant through an appropriate feedback law and, further,  the CBF takes on positive values over the unsafe states.

In this study, we focus on the synthesis of multiple ``min-max''  CBFs for switched affine systems, which are formed by taking pointwise minimum/maximum over a set of affine functions. Specifically, we consider switched affine systems of the form $\dot{x}(t) = A_l x(t) + b_l $.  The control input takes the form of a continuous-time switching signal that selects the mode based on full-state feedback.  We use ideas from non-smooth analysis to formulate mathematical conditions for CBFs. We show that given piecewise min/max CBFs, we can extract a state-based feedback in the form of a mapping from states to sets of possible modes that guarantee safety of the closed loop system. We also extend our previous work on synthesis of piecewise affine control Lyapunov functions for switched affine systems to provide an algorithm for synthesizing CBFs from the description of the switched system~\cite{berger2022learning,Kamali+Berger+Sankaranarayanan/2025/Polyhedral}.
It is well known that synthesizing polyhedral Lyapunov functions is already a hard problem~\cite{berger2022learning}. Extending this to polyhedral CBFs introduces additional challenges, as the functions must satisfy more constraints to ensure safety. Our work combines the min- and max- barrier functions to establish a single min-max CBF.

Our proposed approach is based on the counterexample-guided inductive synthesis (CEGIS) framework~\cite{Solar-Lezama/2008/Program}. This approach is a powerful and widely-used method for synthesizing Lyapunov and barrier functions \cite{abate2021fossil,ahmed2020automated,berger2022learning,berger2023counterexample,chen2021learning,ravanbakhsh2017class}. Based on the CEGIS approach in our study, the algorithm alternates between a verifier and a learner, iteratively refining candidates until a valid CBF is found or the tree search reaches a predefined expansion limit. 

\noindent\textit{Organization:}  Section \ref{sec:Related-work} reviews related work, and Section \ref{sec:Notation} introduces the notations used throughout the paper. Section \ref{sec:Problem-Statement} presents the preliminaries and the problem statement. In Section \ref{sec:Polyhedral-CBF}, we present polyhedral control barrier certificates and pointwise minimum control barriers. The branch-and-bound tree search algorithm is detailed in Section \ref{sec:Algorithm}. Section \ref{sec:Multiple-barriers} explains the Min-Max multiple barriers. Finally, we evaluate our approach using several examples, including a DC-DC converter, a 2D numerical example, a 3D multi-agent system, and a 6D car velocity system, in Section \ref{sec:Examples}. 

\noindent\textit{Appendices:}We have placed the proof of termination of Algorithm~\ref{alg:tree-search-explore} (Cf. Section~\ref{sec:Algorithm}) in Appendix~\ref{app:termination-analysis}. The details of the so-called M-trick for encoding disjunctive constraints as mixed-integer linear programs are in Appendix~\ref{app:milp=encoding}. Finally, Appendix~\ref{app:numerical-tables}  details the CBFs synthesized for our empirical evaluation benchmarks (Cf. Section~\ref{sec:Examples}).

\subsection{Related Work}\label{sec:Related-work}

\noindent\textit{Control Barrier Functions:}  
CBFs were first conceived  by Wieland and Allgower~\cite{wieland2007constructive} as extensions of the notion of barrier functions~\cite{Prajna+Jadbabaie/2004/Safety} to include control inputs, just as control Lyapunov functions~\cite{sontag1983lyapunov,Sontag/1989/Universal} extend the classic notion of a Lyapunov function. Ames et al. in \cite{ames2014control} expanded the definition and introduced the notion of safety filters, that have made important contributions towards assured autonomous systems. Since then, the concept has been extended to address high-degree systems \cite{9516971,9777251}, uncertain systems \cite{wabersich2023data}, and systems with actuation constraints \cite{ames2019control}. Notably, many of these works focus on smooth, single CBFs.

\noindent\textit{Nonsmooth Barrier/Lyapunov Functions:} Nonsmooth CBFs were first proposed by Glotfelter et al. in \cite{glotfelter2017nonsmooth}, where the authors introduced the use of min and max operators to define CBFs for multi-agent systems. Subsequent works extended the concept of nonsmooth barrier functions considerably and investigated the ability to form Boolean combinations of CBFs \cite{ong2023nonsmooth,glotfelter2018boolean,glotfelter2019hybrid}. This paper provides a stronger condition for non-smooth CBFs for switched affine systems and provides algorithms for their synthesis.  


\noindent\textit{Sum Of Squares (SOS) Approaches:} The use of Sum-Of-Squares approaches for CBF synthesis has been investigaged by many, including the recent work of Clark~\cite{clark2024semi}. These approaches have studied the problem of synthesizing CBFs for nonlinear systems with polynomial dynamics.  However, many existing approaches are restricted to control affine systems and result in bilinear optimization problems that are hard to solve.
Here, we focus on an entirely different setup that is characterized by switched affine dynamics and nonsmooth CBFs. Our approach is inspired by a branch-and-bound search with a cutting plane argument to enforce termination. While our approach also suffers from the high complexity of computing CBFs,  we synthesize multiple low complexity CBFs and combine them to create  CBFs with larger control invariant region. This idea of combining simple barriers to create a larger control invariant set was recently explored by Wajid and Sankaranarayanan for polynomial CBFs using SOS approaches that are very different and complementary to the ideas in this paper~\cite{Wajid+Sankaranarayanan/2025/Successive}.

\noindent\textit{Neural CBFs:} Neural networks have also been utilized for constructing CBFs and CLFs. For example, in \cite{abate2021fossil}, the authors introduced FOSSIL, a software tool for synthesizing barrier and Lyapunov functions. Their approach employed neural networks during the learner phase of the CEGIS method to find candidate functions. Poonawala et al. in \cite{poonawalastability} has utilized single-hidden-layer neural network to synthesize a CLF with ReLU NN for a system with single-layer ReLU NN model. Their approach converts the NN to a PWA function and  utilizes nonsmooth analysis to synthesis Lyapunov function and feedback controller.  

\noindent\textit{CEGIS Approaches:} Our approach belongs to a class of methods introduced in the formal methods community that are termed ``counterexample-guided inductive synthesis''~\cite{Solar-Lezama/2008/Program}, wherein the set of  remaining candidate solutions are  refined at each step by choosing a candidate and checking whether it is a valid solution. In our setup, the candidate solutions are the coefficients of the desired CBF.
Ravanbakhsh et al.~\cite{Ravanbakhsh+Sankaranarayanan/2018/Learning} use a SOS programming approach to synthesize CLFs.  Their approach employed CEGIS to address feasibility problems, utilizing an SMT solver to construct polynomial barrier and Lyapunov functions and design controllers. A similar approach was proposed in \cite{zhang2023efficient}, where the authors developed an algorithm to synthesize and verify polynomial CBFs by solving SoS and linear inequalities.

Our work builds on previous ideas used to synthesize CLFs using CEGIS. Berger et al. \cite{berger2022learning} introduced a CEGIS-based method to synthesize polyhedral Lyapunov functions. Their work noted the hardness of proving stability using polyhedral Lyapunov functions.
In our recent work~\cite{Kamali+Berger+Sankaranarayanan/2025/Polyhedral}, we extended this idea to construct polyhedral CLFs for switched affine systems using CEGIS and mixed-integer linear programming (MILP). The key differences in this work include extensions to control barrier functions and combining min/max CBFs. Note that the idea of combining multiple functions is unique to the case of control barriers.

\subsection{Notation}\label{sec:Notation}
Let $\Real_{\geq 0}$ denote the set of non-negative real numbers. For a vector $c \in \Real^m$, $c^t$ denotes the transpose of vector $c$. We will denote the set of natural numbers $\{1,2,3,\cdots\}$ as $\Nat$. For $m \in \Nat$, let $[m] = \{ 1, \ldots, m \}$. And $\operatorname{co}(V)$ shows the convex hull of the set $V$. 

\section{Problem Statement}\label{sec:Problem-Statement}

\begin{defn}[Switched Affine System]\label{def:switched-affine-system}
A switched affine system $\Pi$ with $m \in \Nat$ modes is specified by a set of tuples $\setof{ (A_1, b_1), \ldots, (A_m, b_m) }$
wherein, the dynamics for mode $l \in [m]$ are given by
\begin{equation}
    \dot{x}(t) = A_l x(t) + b_l \,. \label{eq:system_dynamic}
\end{equation}
 Here, $x(t) \in \mathbb{R}^n$ is the state, $l \in [m]$ denotes the mode, $A_l \in \mathbb{R}^{n \times n}$ and $b_l \in \mathbb{R}^n$ for all $l \in [m]$. The initial state $x(0)$ belongs to the initial set $X_0 \subset \mathbb{R}^n$. 
\end{defn}

\begin{figure}[t]
\begin{center}
\includegraphics[width=0.6\columnwidth]{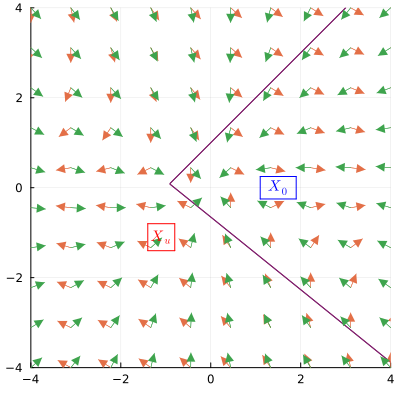}
\end{center}
\caption{Quiver plot showing the two dynamical modes from Ex.~\ref{Ex:Running-Ex-1}, the initial set 
$X_0$ and unsafe set $X_u$. The solid line shows the boundary of the control invariant region defined by the polyhedral CBF.}\label{fig:quiver-running-example}
\end{figure}
\begin{exmp}\label{Ex:Running-Ex-1}
Consider a switched affine system over $\Real^2$ given by two matrices $(A_l, b_l)$ for $l \in [2]$
given as follows:
\[ A_1 = \begin{bmatrix} 
1 & 1 \\ 
0 & -1 \\ 
\end{bmatrix}, \; \text{and}\; A_2 = \begin{bmatrix} 
-1 & 0.1 \\ 
0.2 & -1 \\ 
\end{bmatrix} \,,\]
with $b_1 = b_2 = [0, 0]^t$. The initial set $X_0$ is given by $[1.1, 1.9] \times [-0.25, 0.25]$. Figure~\ref{fig:quiver-running-example} shows the quiver plot.
\end{exmp}

Given a switched affine system $\Pi$, a state-based switching rule  is a set-valued map $\sigma: \Real^n \rightrightarrows [m]$, wherein for any state $x \in \Real^n$, $\sigma(x) \subseteq [m]$ is the set of all the modes that are active for a state $x$.

A differential inclusion over $x \in \Real^n$ is defined as 
\begin{equation}
    \dot{x}(t) \in F(x(t)) \,, \label{eq:differential_inclusion1}
\end{equation}
wherein $F : \mathbb{R}^n \rightrightarrows \Real^n$  is a set-valued map. A solution to a differential 
inclusion (in the Carath\'eodory sense) is a map $x: [0, T) \to \mathbb{R}^n$ for some time $T > 0$ such that 
(a) $x$ is absolutely continuous and (b) $\dot{x}(t) \in F(x(t))$ holds almost everywhere (using the Lesbegue measure) over 
the interval  $[0, T)$.

For a given switched system $\Pi$ and state-based switching rule $\sigma$, we associate a \emph{set-valued} map $F[\Pi, \sigma]$ that 
is defined as follows:
\[ F[\Pi, \sigma](x) = \mathsf{co}\left\{  A_l x + b_l \ |\ l \in \sigma(x) \right\} \,. \]
If $F[\Pi,\sigma]$ is a locally bounded and upper semicontinuous set-valued map with nonempty, convex, and compact values, then the existence of solutions to the differential inclusion $\dot{x}(t) \in F[\Pi, \sigma](x(t))$  is guaranteed~\cite{cortes2008discontinuous}. 
Based on the dynamic of $F[\Pi, \sigma]$, if the switching feedback law $\sigma$ is defined upper semicontinuous and nonempty, then all conditions required for the existence of solutions will be satisfied. In next section, we will define this switching feedback law $\sigma$ based on the proposed polyhedral control barrier certificate.

\section{Polyhedral Control Barrier Certificates}\label{sec:Polyhedral-CBF}

A polyhedral function \(B: \mathbb{R}^n \rightarrow \mathbb{R}\) is defined by a piecewise maximum of a finite number of affine terms:  
\begin{equation}
    B(x) = \max_{i=1}^k c_i^t x - d_i \,,\label{eq:polyhedral-function}
\end{equation}
wherein \(k \in \mathbb{N}\) denotes the number of the pieces, $c_i \in \mathbb{R}^n$ and \(d_i \in \mathbb{R}\). In general, polyhedral functions are Lipschitz continuous but not differentiable.  We would like to use polyhedral barrier certificates to establish controlled invariant sets for switched affine systems. 

Let $\Pi$ be a switched affine system following Def.~\ref{def:switched-affine-system} with initial condition $X_0$,  and $m$ modes whose dynamics are defined by matrices/vectors $\{(A_1, b_1), \ldots, (A_m, b_m)\}$. Let $X_u \subseteq \Real^n$ be a closed set of \emph{unsafe states} that we do not wish to reach. We assume that $X_0 \cap X_u =\emptyset$. 

\begin{defn}[Polyhedral Control Barrier Candidate]\label{def:polyhedral-cbf}
A polyhedral function $B: \mathbb{R}^n \to \mathbb{R} $ given by \eqref{eq:polyhedral-function} is said to be a control barrier candidate for $\Pi$ iff the following conditions hold: 
\begin{description}
    \item[(C1)]  The barrier function must be negative over the initial set of states: $\forall x \in X_0,\ B(x) < 0$, 
    \item[(C2)] The barrier function must be positive over the unsafe states:  $\forall x \in X_u, B(x) > 0$, 
    \item[(C3)] For every state $x$ such that $B(x) \leq 0$, there must exist a dynamical mode $l \in [m]$ such that every piece of the barrier function that is maximized at $x$ must satisfy a ``decrease condition'':
    \[  \begin{array}{l}
    \forall x \in \Real^n, \exists\ l \in [m],\ \forall i \in [k],\\ 
    \ \ \ \ \ \begin{array}{l} \left( 
    B(x) \leq 0\ \land\ c_i^t x - d_i = B(x) \right) \ 
     \Rightarrow\\
     \ \ \ \ \ c_i^t (A_l x + b_l) < - \lambda (c_i^t x - d_i)\, \end{array}  \end{array}\]
    where $\lambda \in \mathbb{R}$ is a constant value. 
\end{description}
\end{defn}

Condition \textbf{(C3)} is analogous to the ``exponential barrier condition'' $\dot{B}(x)  <  -\lambda B(x)$ which would be valid if $B$ were a differentiable function \cite{kong2013exponential}. Note also that we place no condition on points where $B(x) > 0$. We prove the existence of a  switching law $\sigma$ that satisfies the following properties:  (\textbf{E1}) $\sigma(x)$ is an upper semicontinuous set-valued map;  (\textbf{E2}) $\sigma$ induces the decrease of $B(x)$ at $x$ whenever $B(x) < 0$;  \textbf{(E3)} using $\sigma$ as a feedback law results in the set $\{ x \ |\ B(x) < 0 \}$ being a positive invariant set.  Therefore, we conclude that  any trajectory of the closed-loop system $\dot{x}(t) \in F[\Pi, \sigma](x(t))$ starting at a state $x(0)$ wherein $B(x(0)) < 0$ will be safe: i.e,  for all $t \geq 0$, $x(t) \not\in X_u$.


\subsection{Switching Feedback Law}
\begin{defn}[Upper Semicontinuity]\cite{Aubin2009}\label{def:upper-semi}
    A set-valued map $\sigma: X \rightrightarrows [m] $ is upper semicontinuous at $x \in X$ if and only if for any neighborhood $\mathscr{U}$ of $\sigma(x)$, there exists $r>0$ such that for all $y \in B_X(x,r), ~ \sigma(y) \subset \mathscr{U}.$
\end{defn}
\begin{prop}\cite{Aubin2009}\label{prop:Upper-semicontinuous-closed}
    If $\text{Dom}(\sigma)$ is closed, then $\sigma$ is upper semicontinuous if and only if $\sigma^{-1}(l)$ is closed for every closed subset in $\text{Dom}(\sigma)$ for every $l\in[m]$.
\end{prop}

Based on the Definition \ref{def:upper-semi} and Proposition \ref{prop:Upper-semicontinuous-closed}, we formalize both conditions \textbf{\textit{(E1)}} and \textbf{\textit{(E2)}} by defing the switching law with the closure property as follow. 
\begin{defn}\label{def:closure-prop}
    $\sigma:\Real^n\rightrightarrows[m]$ has the closure property if
    \begin{description}
        \item[(G1)] $\sigma(x)$ is  nonempty for all $x\in\Real^n$, 
        \item[(G2)] $\sigma^{-1}(l) = \{ x \ |\ l \in \sigma(x) \} $ is closed for all $l\in[m]$, 
        \item[(G3)] for all $x\in\Real^n$, if $l \in \sigma(x)$ then 
        \begin{align}
        (\forall i \in [k]), ~~~ B(x) & = c_i^t x - d_i \leq 0 \Rightarrow \\\nonumber
        & c_i^t (A_l x + b_l)  < -\lambda (c_i^t x - d_i).
    \end{align}
    \end{description}
\end{defn}
We will define the switching law that has the closure property by idea of defining a \emph{merit function}  as follows. 
\begin{defn}[Associated Switching Rule]\label{def:associated-switching-rule}
Given a polyhedral function $B: \mathbb{R}^n \to \mathbb{R}$ that satisfies the conditions \textbf{(C1)-(C3)} in Def.~\ref{def:polyhedral-cbf}, we define the \emph{associated switching rule} $\sigma: \Real^n \rightrightarrows [m]$. 
Let us consider an arbitrary point $x \in\Real^n$.

\noindent\textbf{Case-1:} If $B(x) \leq 0$, we define $\sigma(x)$ as
\begin{equation}
    \sigma(x) = \operatorname{argmax}_{l \in [m]}  \mathscr{M}(x,l),\quad \text{wherein}\ 
\end{equation}
\begin{equation}
    \mathscr{M}(x,l) = \min\limits_{i \in [k]}  \mathscr{M}_i(x,l),\quad \text{with}\ 
\end{equation}
\begin{equation}\label{eq:merit-i}
    \mathscr{M}_i(x,l) =  \max\limits_{\tau \geq 0}\{ \tau | \tau \hat{\varphi}_i + \varphi_i + \lambda \tau \varphi_i - \varphi \leq - \tau^2  \} 
\end{equation}
where $\hat{\varphi}_i = c_i^t (A_l x + b_l)$, $\varphi_i = c_i^t x - d_i$, and $\varphi = \max\limits_{i \in [k]} c_i^t x - d_i$. The function 
$\mathscr{M}(x, l)$ will be denoted the \emph{merit function} at state $x$ for choice of dynamics $l$.

\noindent\textbf{Case-2:} For each $x$ with $B(x) > 0$, we define
\begin{equation}
    \sigma(x) = \{\sigma(x')\ |\ x' = \operatorname{argmin}\limits_{z \in \mathbb{R}^n} \{\| x-z\| \; | B(z) \leq 0\}\}. \label{eq:sigma-for-positive-B}
\end{equation}
In other words, we will project the point $x$ onto the set $ \{ x \ | \ B(x) \leq 0 \}$ and reuse the definition from case-1.
\end{defn}
\begin{prop}\label{prop:merit-function}
    The proposed merit function satisfies the following properties: 
    \begin{enumerate}
         \item $ \mathscr{M}(x,l) \geq 0$ for all $x$ and $l$;
        \item $ \mathscr{M}(x,l) > 0$ if only if $~\forall i \in [k], ~ \left( B(x) = c_i^t x - d_i \leq 0 \right)$ $\Rightarrow c_i^t(A_l x + b_l) < -\lambda (c_i^t x - d_i)$;
        \item $ \mathscr{M}(x,l)$ is continuous w.r.t $x$. 
    \end{enumerate}
\end{prop}
\begin{pf}
        (1) If $\tau = 0$, then the inequality in \eqref{eq:merit-i} holds, since $\varphi_i \leq \varphi$ by definition. Therefore, $\mathscr{M}_i(x, l) \geq 0$ for all $x, l$. Hence $\mathscr{M}(x, l)= \min_{i \in [k]} \mathscr{M}_i(x,l) \geq 0$.
        
       (2) ($\Longleftarrow$) Assume the decrease condition holds, i.e., $\forall i \in [k], ~\varphi = \varphi_i \leq 0 \implies \hat{\varphi}_i < -\lambda \varphi_i$.  Since $\varphi = \max_{i \in [k]} \varphi_i$ we have either: case (i) $\varphi = \varphi_i$ and case (ii) $\varphi_i < \varphi$.
       \begin{compactdesc}
       \item[Case (i):] Considering $\varphi = \varphi_i$, the inequality in \eqref{eq:merit-i} can be rewritten as $\tau \hat{\varphi}_i + \lambda \tau \varphi_i \leq -\tau^2$. Considering this and the decrease assumption, $\hat{\varphi}_i < -\lambda \varphi_i $, the inequality in \eqref{eq:merit-i} holds for $\tau \in [0, -(\hat{\varphi}_i+ \lambda \varphi_i)]$. Based on the decrease assumption, $-(\hat{\varphi}_i+ \lambda \varphi_i) > 0$ which results in for all $i\in [k]$, $\mathscr{M}_i(x,l) = \max\{0, -(\hat{\varphi}_i+ \lambda \varphi_i)\} = -(\hat{\varphi}_i+ \lambda \varphi_i)$ which is a positive value.
       \item[Case (ii):] Let's define $\vartheta(\tau) = \tau \hat{\varphi}_i + \varphi_i + \lambda \tau \varphi_i - \varphi + \tau^2$. The function $\vartheta(\tau)$ is an upward quadratic function with roots $\bar{\tau}_{1,2} = \frac{-(\hat{\varphi_i}+\lambda \varphi_i) \pm \sqrt{(\hat{\varphi_i}+\lambda \varphi_i)^2-4(\varphi_i - \varphi)}}{2}$. Since $\varphi_i < \varphi$, then $\bar{\tau}_{i_{1,2}}$ are real values. Moreover, $\vartheta(0) = \varphi_i - \varphi < 0$. Considering all, it can be easily concluded $\vartheta(\tau) \leq 0$ for all $\tau \in [0, \bar{\tau}_{i_1}]$ where $\bar{\tau}_{i_1} = \frac{-(\hat{\varphi_i}+\lambda \varphi_i) + \sqrt{(\hat{\varphi_i}+\lambda \varphi_i)^2-4(\varphi_i - \varphi)}}{2}$ which results in the inequality in \eqref{eq:merit-i} holds for $\tau \in [0, \bar{\tau}_{i_1}]$. Therefore, $\mathscr{M}_i(x,l) = \bar{\tau}_{i_1} > 0$ and hence from two cases $\mathscr{M}(x,l) $ is greater than 0.
       \end{compactdesc}
       ($\Longrightarrow$) We prove this by contradiction. Assume that the decrease condition does not hold, i.e., there exists $i \in [k]$ such that $\varphi = \varphi_i \leq 0 \implies \hat{\varphi}_i \geq -\lambda \varphi_i$. So, we can write $\exists i \in [k]$ such that $\vartheta(\tau) = \tau(\hat{\varphi}_i + \lambda \varphi_i) + \tau^2$ which is less than or equal zero just at $\tau = 0 $. So, there exists $i \in [k]$ such that $\mathscr{M}_i(x,l) = 0$ and hence $\mathscr{M}(x,l) = 0$. 
       
       (3) If $\varphi_i = \varphi$, then $\mathscr{M}_i(x, l) = -(\hat{\varphi}_i + \lambda \varphi_i) = - c_i^t (A_l x + b_l) - \lambda (c_i^t x - d_i)$ which is continuous map w.r.t. $x$. And if $\varphi_i < \varphi$, then $\mathscr{M}_i(x, l) = \bar{\tau}_{i_1}$ which is a continuous function w.r.t. $x$. Thus, the merit function $\mathscr{M}(x,l) = \min\limits_{i \in [k]} \mathscr{M}_i(x,l)$ is continuous.  
\end{pf}
\begin{thm}\label{thm:closure-property}
    The proposed switching law $\sigma$ in Definition \ref{def:associated-switching-rule} has the closure property. 
\end{thm}
\begin{pf}
    \textbf{\textit{(G1)}} and \textbf{\textit{(G3)}} are direct consequence of condition \textbf{(C3)} from Def.~\ref{def:polyhedral-cbf}, the definition of the associated switching rule in Def.~\ref{def:associated-switching-rule}, and its properties in Prop.~\ref{prop:merit-function}. To prove \textbf{\textit{(G2)}}, consider $B(x)$ is a CBF. 
    Let $l\in[m]$.
    For any $x\in\Real^n$ such that $B(x)\leq0$, it holds that $l\in\sigma(x)$ if and only if $\mathscr{M}(x,l)-\max_{l'\in[m]}\mathscr{M}(x,l')\geq0$.
    Hence, $\{x \mid B(x)\leq0,\,l\in\sigma(x)\}$ is the pre-image of $[0,\infty)$ by the continuous function $\mathscr{M}(x,l)-\max_{l'\in[m]}\mathscr{M}(x,l')$.
    Hence, it is closed since the pre-image of a closed set by a continuous function is closed. 
    On the other hand, from \eqref{eq:sigma-for-positive-B}, for all $x\in\Real^n$ such that $B(x)\geq0$, $l\in\sigma(x)$ if and only if \textcolor{red}{$\min_{\{z\mid B(z)\leq0,l\in\sigma(z)\}}\lVert x-z\rVert - \min_{\{ z\mid B(z)\leq0 \}}\lVert x-z\rVert \leq 0$.}
    Hence, for the some reason as above, $\{x \mid B(x)\geq0,\,l\in\sigma(x)\}$ is closed.
\end{pf}

\begin{thm} \label{thm:F-upper-semi}
    If $\sigma$ be upper semicontinuous, then $F[\Pi, \sigma]$ is upper semicontinuous. 
\end{thm}
\begin{pf}
    Let $x\in\Real^n$.
    Let $(x_n)_{n\in\mathbb{N}}$ be a sequence converging to $x$.
    Assume $\sigma$ is upper semicontinuous. So, 
    \[\exists N \in \mathbb{N}\; s.t. \; \forall n\geq N,\; \sigma(x_n) \subseteq \sigma(x)\]
    which results in 
    \begin{align*}
        &\forall \varepsilon>0 \;  \exists N \in \mathbb{N}\; s.t. \; \forall n\geq N,\\
        & \{A_lx_n + b_l\;| l \in \sigma(x_n)\} \subseteq \{A_lx + b_l\;| l \in \sigma(x)\} + \varepsilon \mathscr{B}
    \end{align*}
    where $\mathscr{B}$ is a unit ball. Applying the convex hull, 
    \begin{align*}
        &\forall \varepsilon>0 \;  \exists N \in \mathbb{N}\; s.t. \; \forall n\geq N,\\
        & \operatorname{co}\{A_lx_n + b_l\;| l \in \sigma(x_n)\} \subseteq \operatorname{co}\{A_lx + b_l\;| l \in \sigma(x)\} + \varepsilon \mathscr{B}
    \end{align*}
    which is equivalent to 
    \begin{align*}
        & \forall \varepsilon > 0, \exists N \in \mathbb{N}\; s.t. \; \forall n\geq N,\\
        & F[\Pi, \sigma](x_n) \subseteq F[\Pi, \sigma](x) + \varepsilon \mathscr{B}
    \end{align*}
    that means $F[\Pi, \sigma]$ is upper semicontinuous. 
\end{pf}
The following result is a direct consequence of Definition \ref{def:closure-prop}, and Theorems \ref{thm:closure-property} and \ref{thm:F-upper-semi}. 
\begin{thm}
    If the switching law $\sigma$ has the closure property,  all trajectories starting from $X_0$ are complete.
\end{thm}

\label{subsecion-soundness}
\section{Soundness of Polyhedral Barrier Certificate}


We will now prove the soundness of polyhedral control barrier functions for a switched affine system $\Pi$ with polyhedral barrier function
$B$ and associated switching rule $\sigma$. First, we recall some facts from nonsmooth analysis that will be helpful for establishing the results. Then, we characterize the weak Lie derivative of $B$ w.r.t the differential inclusion $F[\Pi, \sigma](x(t))$ and show the soundness. For convenience, we will write $F = F[\Pi, \sigma]$ throughout this section. \\

\subsection{Nonsmooth Analysis}
\label{section:Nonsmooth-Invariance for Differential Inclusions}
We will now recall some facts from nonsmooth analysis that will be helpful in establishing the key results in this paper. We refer the reader to the survey by Cortes~\cite{cortes2008discontinuous} and the textbooks of Clarke~\cite{Clarke/1983/Optimization} and Lakshmikantham et al.~\cite{Lakshmikantham+Leela/1969/Differential} for detailed definitions and proofs. 
Let $\dot{x}(t) \in F(x(t))$ be a differential inclusion, wherein $F$ is locally bounded and upper semicontinuous. We will assume that all differential inclusions in this section  are locally bounded and upper semicontinuous, thus guaranteeing the existence of solutions. Let $g(x) = \max_{i=1}^k g_i(x)$ be a piecewise maximum of finitely many smooth (infinitely differentiable) functions $g_1, \ldots, g_k$.
We recall Prop. 2.3.12 in \cite{clarke1990optimization}:
\begin{thm}\label{thm:generalized-gradient-property}
For all $x\in\Real^n$, it holds that $\partial g(x) \subseteq \conv \left(\{ \nabla g_j(x)\ |\ j \in [k], g_j(x) = g(x) \}\right)$.
\end{thm}
\begin{defn}[Weak Set-Valued Lie Derivative]
The weak Lie derivative of a function $g$ w.r.t. the differential inclusion $F$ is given by 
\begin{equation}
\label{eq:weak-set-valued-derivative}
    \mathcal{L}_F^w g(x) = \{ \theta \cdot \xi\  |\ \theta \in F(x),\,  \xi \in \partial g(x) \} \,,
\end{equation}
wherein $\partial g(x)$ is the Clarke generalized gradient of $g$ at $x$, and $a \cdot b$ denotes the inner product of vectors
$a, b$.
\end{defn}
The following lemma is useful for computing $\partial g$ wherein $g = \max_{i=1}^k g_i$ for Lipschitz continuous functions $g_i$.
\begin{lem}\label{Lem:useful-lemma}
Let $g(x)  = \max_{i=1}^k g_i(x)$, wherein $g_i(x)$ are smooth functions and $F(x) = \conv\{ f_1(x), \ldots, f_m(x) \}$. 
\[ \mathcal{L}_F^w g(x) \subseteq \conv \{ f_i(x) \cdot \nabla g_j(x)\ |\ i \in [m], j \in [k], g_j(x) = g(x) \} \,.\]
\end{lem}
\begin{pf}
From Eq.~\eqref{eq:weak-set-valued-derivative}, we have that each element $a$ of $\mathcal{L}_F^w$ is an inner product of the form 
$\theta \cdot \xi$ wherein $\theta \in F(x)$ and $\xi \in \partial g(x)$. By assumption, we may write 
$\theta = \sum_{i=1}^m \lambda_i f_i$, wherein $\lambda_i \geq 0$ for $i \in [m]$ and $\sum_{i \in [m]} \lambda_i = 1$.
From Theorem~\ref{thm:generalized-gradient-property}, we note that $\partial g(x) \subseteq \conv \left(\{ \nabla g_j(x)\ |\ j \in [k], g_j(x) = g(x) \}\right)$. Therefore, consider the set $G(x) = \{ \nabla g_j(x)\ |\ j \in [k], g_j(x) = g(x) \} $.  We have 
$\mathcal{L}_F^w g(x) \subseteq L(x)$ wherein $L(x) = \{ \theta \cdot \xi \ |\ \theta \in F(x), \xi \in G(x) \}$. 
Each $a \in L(x)$ can be written as 
$a = (\sum_{i=1}^m \lambda_i f_i) \cdot (\sum_{j=1}^k \gamma_j g_j ) = \sum_{i=1}^m \sum_{j=1}^k \lambda_i \gamma_j (f_i \cdot g_j)$,
wherein $\gamma_j \geq 0$, and $\sum_{j\in [k]} \gamma_j = 1$. Thus,  $\lambda_i \gamma_j \geq 0$ and $\sum_{i \in [m]} \sum{j\in [k]} \lambda_i \gamma_j = 1$, yielding
$a \in \conv \{ f_i(x) \cdot g_j(x)\ |\ i \in [m], j \in [k] \}$.
\end{pf}
\begin{thm}\label{thm:derivative-of-h}
Let $x: [0, T) \mapsto \Real^n$ be an absolutely continuous function that is a solution to the differential inclusion $\dot{x}(t) \in F(x(t)) $, and $g$ be a locally Lipschitz continuous function. The inclusion holds almost everywhere in the interval $[0, T)$:
$ \frac{d}{dt}g(x(t)) \in \mathcal{L}_F^w g(x(t))\,.$
\end{thm}
\begin{pf}
Proof follows from Lemma 1 of~\cite{bacciotti1999stability}, Proposition 2.2.2 of Clarke's book~\cite{Clarke/1983/Optimization}, and Remark 2 of \cite{glotfelter2017nonsmooth}.
\end{pf}

For a scalar $\lambda \in \Real$, we define the function $\alpha: \Real \rightarrow \Real$:
\begin{equation}\label{eq:alpha-fun-def}
    \alpha(s) = \begin{cases} 
- \lambda s & s \leq 0 \\ 
\infty & \text{otherwise} 
\end{cases} 
\end{equation}
Since, $\alpha$ is a Lipschitz continuous function,  the ODE $\dot{s} = \alpha(s)$ with initial condition 
$s(0) = s_0 \in \Real_{\leq 0}$ has a unique solution $s(t) = e^{-\lambda t} s_0$ when $s_0 \leq 0$. Therefore, 
\begin{equation}\label{eq:alpha-fun-ode-sol}
s(t) \leq \begin{cases}
    e^{-\lambda t} s_0 & \text{if}\ s_0 \leq 0 \\ 
    \infty & \text{otherwise.}
\end{cases}
\end{equation}

The following useful ``comparison principle'' follows from Theorem 1.10.2 of Lakshmikantham and Leela~\cite{Lakshmikantham+Leela/1969/Differential}.
\begin{thm}\label{Thm:useful-comparison-principle}
Let $x: [0, T) \rightarrow \Real^n$ be an absolutely continuous function that is a solution of the differential inclusion 
$\dot{x}(t) \in F(x(t))$ and $g: \Real^n \rightarrow \Real$ 
be a Lipschitz continuous function such that for almost every $t \in [0,T)$, 
\[ \frac{d}{dt} g(x(t))  ~<~\alpha( g(x(t)) )\,,\ \text{and}\  g(x(0)) \leq 0 \] 
Thus, $g(x(t)) ~\leq~ e^{-\lambda t} g(x(0)) \leq 0$ for all $t \in [0, T)$.
\end{thm}

\subsection{Soundness of Polyhedral Barrier Certificate}\label{sec:polyhedral-barrier-sound}
Now we prove that any trajectory $x: [0, T) \rightarrow \Real^n$ that is a
solution to the differential inclusion $ \dot{x}(t) \in F(x(t))$ and initial condition $x(0)$ satisfying $B(x(0)) < 0$ where $B(x(t))$ is a polyhedral barrier certificate, will 
satisfies $B(x(t)) ~{<}~ 0$ for all time $t \in [0, T)$. 

\begin{lem}\label{Lem:weak-lie-deriv-barrier}
For all $x \in \Real^n$ the following inclusion holds: 
\[ \mathcal{L}_F^w B(x) \subseteq  \conv \left\{ c_i^t (A_l x + b_l) \ \left|\ \begin{array}{c}
i \in [k], c_i^t x - d_i = B(x) \\ 
l \in \sigma(x)
\end{array} \right. \right\} \,.\]
\end{lem}
\begin{pf}
First, $F[\Pi, \sigma](x) = \conv \{ A_l x + b_l\ |\ l \in \sigma(x) \}$.  Next, $B = \max_{i=1}^k B_i$, wherein 
each $B_i(x) = c_i^t x - d_i$ is a smooth function of $x$. Applying Lemma~\ref{Lem:useful-lemma}, we obtain 
\[ \mathcal{L}_F^w B(x) \subseteq \conv \left\{ f_j \cdot \grad B_i \ \left|\ \begin{array}{c}
i \in [k], j \in [m], \\ 
f_j \in F(x),\ B_i(x) = B(x) \end{array} \right. \right\} \,.\]
\end{pf}

We now connect $\mathcal{L}_F^w B(x)$ with $\alpha(B(x))$, wherein $\alpha$ is the function defined
in ~\eqref{eq:alpha-fun-def}.

\begin{lem}\label{Lem:weak-deriv-inclusion}
For all $x \in \Real^n$, $\sup \left( \mathcal{L}_F^w B(x) \right) \leq \alpha(B(x)) $.
\end{lem}
\begin{pf}
\textbf{Case-1:} Assume $B(x) \leq 0$. From Lemma~\ref{Lem:weak-lie-deriv-barrier}, whenever $B(x) \leq 0$
\[ \mathcal{L}_F^w B(x) \subseteq  \conv \underset{S(x)}{\underbrace{\left\{ c_i^t (A_l x + b_l) \ \left|\ \begin{array}{c}
i \in [k], c_i^t x - d_i = B(x) \\ 
l \in \sigma(x)
\end{array} \right. \right\}}} \,. \]
However, from the definitions of $B$ and $\sigma$, we note that for all $l \in \sigma(x)$, and
for all $i \in [k]$ such that $c_i^t x - d_i = B(x)$, we have 
$ c_i^t (A_l x  +b_l) < \lambda (c_i^t x - d_i) $.
Therefore, for all $a \in S(x)$, we conclude that $a ~{<} -\lambda B(x) = \alpha(B(x))$.
Hence,
$\mathsf{sup}(\mathcal{L}_F^w B(x)) \leq \mathsf{sup}(S(x)) ~{<}~ \alpha(B(x))$.

\textbf{Case-2:} Assume $B(x) > 0$. Again, from Lemma~\ref{Lem:weak-lie-deriv-barrier} 
$ \mathcal{L}_F^w B(x) \subseteq  S(x) $.
Hence, each value of $\mathcal{L}_F^w B(x)$ is less or equal to infinity. So, when $B(x)>0$, we have $\mathsf{sup}(\mathcal{L}_F^w B(x)) \leq \mathsf{sup}(S(x)) \leq \infty$. Thus, by comparing with \eqref{eq:alpha-fun-def},  $\mathsf{sup}(\mathcal{L}_F^w B(x)) \leq \alpha(B(x))$ for $B(x)>0$.
\end{pf}


We can now establish the soundness result that follows directly from  Theorem~\ref{thm:derivative-of-h} and Lemma~\ref{Lem:weak-deriv-inclusion} with the comparison principle
established in Theorem~\ref{Thm:useful-comparison-principle}.
\begin{thm}
Let $B: \mathbb{R}^n \to \mathbb{R}$ be a polyhedral control barrier with the associated map 
$\sigma$. Let $x: [0, T) \rightarrow \Real^n$ be any solution to the differential inclusion 
$\dot{x}(t) \in F[\Pi, \sigma](x(t))$ such that $B(x(0)) \leq 0$. 
For all $t \in [0, T)$, $B(x(t)) ~\leq~ e^{-\lambda t} B(x(0)) \leq 0$.
\end{thm}

\subsection{Min Control Barrier Certificate}\label{sec:min-barriers}
In this section, we consider CBFs that are piecewise minima of a finite number of the affine terms as
\begin{equation}
\label{eq:min-barriers}
    B(x) = \min_{i=1}^k c_i^t x - d_i
\end{equation}

\begin{lem}
\label{lem:Generalized-Clarke-piecewise-min}
    Let $g(x) = \min_{i=1}^k g_i(x)$ for smooth $g_i$.  
    \[ \partial g (x) \subseteq \conv \left(\{ \nabla g_j(x)\ |\ j \in [k], g_j(x) = g(x) \}\right)\,. \]
\end{lem}
\begin{pf}
    The proof follows from Theorem \ref{thm:generalized-gradient-property} and 
    $\min_{i=1}^k g_i(x) = - \max_{i=1}^k (-g_i(x))$. 
\end{pf}
Moreover, the piecewise minimum functions holds the locally Lipschitz property. Considering this and Lemma \ref{lem:Generalized-Clarke-piecewise-min}, all the lemmas and theorems in section~\ref{sec:polyhedral-barrier-sound} can be applied to piecewise minimum functions.
\begin{defn}[Piecewise Min Control Barriers]
    A piecewise minimum function given by (\ref{eq:min-barriers}) is said to be a control barrier certificate for $\Pi$ iff  the conditions \textbf{(C1)}, \textbf{(C2)} (for polyhedral CBFs) and \textbf{($C'3$)} hold for a constant $\lambda \geq 0$:
    
        \textbf{($C'3$)} For every state $x$ such that $B(x) \leq 0$, there must exist a dynamical mode $l \in [m]$ such that every piece of the barrier function that is \underline{minimized} at $x$ must satisfy a ``decrease condition'':
    \[  \begin{array}{l}
    \forall x \in \Real^n, \exists\ l \in [m],\ \forall i \in [k],\\ 
    \ \ \ \ \ \begin{array}{l} \left( 
    B(x) \leq 0\ \land\ c_i^t x - d_i = B(x) \right) \ 
     \Rightarrow\\ 
     \ \ \ \ \ \ \ \ c_i^t (A_l x + b_l) < - \lambda (c_i^t x - d_i)\,. \end{array}  \end{array}\]
\end{defn}

To satisfy the closure property, we define the merit function related to the associated switching rule for the piecewise minimum as
\[\mathscr{M}(x,l) = \min\limits_{i \in [k]} \mathscr{M}_i^{min}(x,l),\quad \text{wherein}, \] 
\begin{equation}
    \mathscr{M}_i^{min}(x,l) =  \max\limits_{\tau \geq 0}\{ \tau | \tau \hat{\varphi}_i - \varphi_i + \lambda \tau \varphi_i + \varphi \leq - \tau^2  \}. 
\end{equation}
In a similar manner to what was demonstrated in Proposition \ref{prop:merit-function} and Theorem \ref{thm:closure-property}, the associated switching rule to this merit function also preserves the closure property.
\begin{rem}
    The piecewise minimum over convex functions is convex, and consequently, the regularity property which results in the inclusion in Theorem \ref{thm:derivative-of-h} holds for strong set-valued Lie derivative \cite{glotfelter2017nonsmooth} as 
    \[\frac{d}{dt}g(x(t)) \in \mathcal{L}_F^s(g(x(t))) ~~~ a.e., wherein, \]
    \[\mathcal{L}_F^s(g(x)) = \{r \in \mathbb{R}| \exists \theta \in F(x), \forall \xi \in \partial g(x): \theta \cdot\xi = r \}. \]
    The strong derivative can provide less conservative results for finding nonsmooth barrier certificate. 
\end{rem}

\section{Branch and Bound Tree Search Algorithm}\label{sec:Algorithm}
In this section, we will provide an algorithmic approach to search for  max (polyhedral) barrier certificates as functions of the form  $B(x) = \max_{i=1}^k c_i^t x - d_i $.   Our algorithm is adapted from our previous works on synthesis of polyhedral Lyapunov and control Lyapunov functions~\cite{berger2022learning,Kamali+Berger+Sankaranarayanan/2025/Polyhedral}. Assume that the coefficients $c_i, d_i$ are bounded by user-input parameter $\gamma$: 
    \begin{equation}\label{eq:cd-bounds}
        \psi_0:= \bigwedge_{i=1}^k \left( -\gamma\ 1_n \leq c_{i} \leq \gamma\ 1_n \ \land\ -\gamma \leq d_i \leq \gamma \right) 
    \end{equation}
$1_n$ refers to the $n\times 1$ vector all of whose entries are $1$.

\begin{defn}[CBF  Synthesis Problem]\label{def:cbf-synthesis-problem}
The polyhedral CBF synthesis problem is as follows:
\begin{itemize}
    \item \textsc{Inputs:} Switched affine system $\Pi$ with  $m$ modes  $(A_i, b_i)$, $i = 1, \ldots, m$, polyhedral initial set $X_0 \not= \emptyset$, polyhedral unsafe set $X_u \not= \emptyset$, \# pieces $k > 0$, constant $\lambda > 0$, parameters $\epsilon > 0$, $\gamma > 0, R_{\min} > 0$. 
    \item \textsc{Output:} Polyhedral CBF $B(x) = \max_{i=1}^k c_i^t x - d_i $ satisfying the conditions in Def.~\ref{def:polyhedral-cbf} 
    or \textsc{Fail}, denoting that no CBF was discovered. 
\end{itemize}
\end{defn}

\begin{algorithm}[t]
\footnotesize 
\DontPrintSemicolon
\KwData{See Def.~\ref{def:cbf-synthesis-problem}}
\KwResult{CBF $B$; or FAIL}
\textbf{Initialize}: root $\alpha_0$ with  $C(\alpha_0) : \psi_0$ (see ~\eqref{eq:cd-bounds}), $W(\alpha_0) = \emptyset$, 
$p \ \leftarrow\ 1$ \;\\
\While{ $\exists$ Unexplored Leaf}{
    $\beta\ \leftarrow\ $  Choose unexplored leaf and mark as explored \;
    
     \lIf{$\left( C(\beta) \text{infeas. \label{nl:termination-criterion}} \ \lor\ \mathsc{ChebyshevRadius}(C(\beta)) \leq R_{\min}\right)$}{\textbf{continue} \;}
    
    $(c_i^{(p)}, d_i^{(p)})_{i \in [k]} \ \leftarrow\ \mathsc{ChooseMVECenter}(C(\beta)) $  \label{nl:choose-candidate}\;
    
     $\textsf{result}\ \leftarrow\ \mathsc{VerifyCandidate}((c_i^{(p)}, d_i^{(p)})_{i \in [k]})$ \label{nl:verify-candidate}\;

    \eIf{$\mathsf{result} = \mathsc{verified}$}{  Return\ CBF $B(x) := \max_{i\in [k]}(c_i^{(p)} x -  d_i^{(p)})$ \label{nl:return-cbf}}
    {
        Create children $\beta_{j, i}$,  for $i \in [k], j \in [m]$  \label{nl:create-children}\; 

        $W(\beta_{j,i}) \ \leftarrow\ W(\beta) \cup \{ (x, j, i ) \}$\; 

         $C(\beta_{j,i})\ \leftarrow\ C(\beta) \ \land\ CW( \{ (x, j, i ) \} ) $ Cf.~\eqref{eq:cw-alpha-constraint} \label{nl:define-c-beta}\; 
        
    }
    
    $p  \ \leftarrow\ p + 1$ \;
}
\Return FAIL \;

\caption{Tree search algorithm pseudocode.}\label{alg:tree-search-explore}
\end{algorithm}

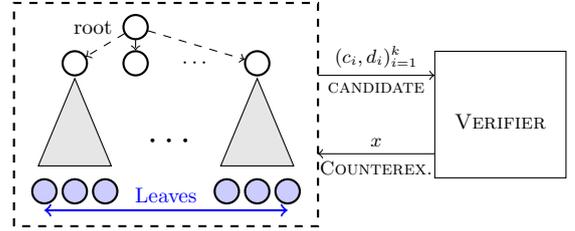
\begin{figure}[t]
\begin{center}


\begin{tikzpicture}[
    scale=0.8, 
    tree node/.style={
        draw, 
        thick,
        circle, 
        minimum size=0.4cm, 
        inner sep=0pt, 
        font=\small 
    },
    leaf node/.style={
        draw, 
        thick,
        fill=blue!20,
        circle, 
        minimum size=0.4cm, 
        inner sep=0pt, 
        font=\small 
    },
    subtree node/.style={
        draw, 
        fill=gray!20, 
        isosceles triangle, 
        minimum width=1.2cm, 
        minimum height=1cm, 
        inner sep=0pt, 
        anchor=south, 
        shape border rotate=90 
    },
    level 1/.style={
        sibling distance=3cm, 
        level distance=1.5cm 
    },
    level 2/.style={
        sibling distance=1.5cm, 
        level distance=1.5cm 
    },
    level 3/.style={
        sibling distance=0.75cm, 
        level distance=1.5cm 
    },
    edge from parent/.style={
        draw, 
        thick, 
        -latex 
    }
]
    \begin{scope}[transform shape]

        
    \draw[dashed, thick] (-2cm, -1.5cm) rectangle (3cm, 2.2cm); 

    \node[tree node] (n0) at (0, 1.8){};
    \node[left of=n0, node distance=0.7cm]{root};
    \node[tree node] (n1) at (-1,  1.2){};
    \node[tree node] (n2) at (0, 1.2){};
    \node(n4) at (1, 1.2){$\cdots$};
    \node[tree node] (n5) at (2, 1.2){};

    \node[subtree node, below of = n1, node distance=1.3cm](st1){};
    \node[subtree node, below of = n5, node distance=1.3cm](st2){};
    \node[below of=n2, xshift=0.6cm, node distance=1.3cm]{\Large $\cdots$};
    \node[below= 0.2cm of st1, xshift=-0.5cm, leaf node ] (l1) {}; 
    \node[below= 0.2cm of st1, xshift=0cm, leaf node ] (l2) {}; 
    \node[below= 0.2cm of st1, xshift=0.5cm, leaf node ] (l3) {}; 
     \node[below= 0.2cm of st2, xshift=-0.5cm, leaf node ] (l4) {}; 
    \node[below= 0.2cm of st2, xshift=0cm, leaf node ] (l5) {}; 
    \node[below= 0.2cm of st2, xshift=0.5cm, leaf node ] (l5) {}; 

    \path[blue, thick, <->] ($ (l1.south) + (0,-0.1cm) $) edge node[above]{Leaves} ($ (l5.south) + (0,-0.1cm) $);

    \draw[->, dashed] (n0) edge(n1) 
        (n0) edge (n2) 
        (n0) edge (n5);

 \node[rectangle, draw=black](verifier) at (6, 0.35){\begin{tabular}{c}
 \\[5pt]\textsc{Verifier}\\[15pt] \end{tabular}};       

\path[->](3, 1) edge node[above]{\footnotesize $(c_i, d_i)_{i=1}^k$} node[below]{\footnotesize \textsc{candidate}} ($ (verifier.west) + (0, 0.65cm) $)
($ (verifier.west) + (0, -0.65cm) $) edge node[above]{\footnotesize $x$}  node[below]{\footnotesize \textsc{Counterex.}} (3, -0.3);
 
\end{scope} 
\end{tikzpicture}
\end{center}
\caption{Branch-and-Bound Tree Search Overview.}\label{fig:alg-overview-fig}
\end{figure}
\noindent\textbf{Overview:} The branch-and-bound tree search algorithm maintains a tree where each node $\alpha$ of the tree has the following information:
(a) a polyhedral constraint $C(\alpha)$ over the unknown coefficients $(c_i, d_i)_{i=1}^k$ and (b) a set of witnesses $W(\alpha): \{ (x, j, i)\ |\ x \in \Real^n, j \in [m], i \in [k] \}$. We will discuss the meaning of each witness and the connection between $W(\alpha)$ and $C(\alpha)$ subsequently.  The pseudocode of the algorithm is provided as Algorithm~\ref{alg:tree-search-explore}. Its operation of the algorithm is illustrated in  Figure~\ref{fig:alg-overview-fig}.
At each step, the algorithm \textbf{proposes a candidate CBF} (line~\ref{nl:choose-candidate}) from the tree which is \textbf{checked by a verifier} (line~\ref{nl:verify-candidate}) that checks conditions \textbf{(C1)-(C3)}. If the check succeeds, then the approach yields a valid CBF (line~\ref{nl:return-cbf}). Otherwise, it yields a counterexample that is used to \textbf{refine the tree} (lines~\ref{nl:create-children}-~\ref{nl:define-c-beta}).  

\noindent\textbf{Tree node:}  The polyhedral constraint $C(\alpha)$ associated with node $\alpha$ describes the set of unexplored CBF candidates $(c_i, d_i)_{i\in [k]}$, while the witnesses $W(\alpha)$ is a set  that contains elements of the form $(x, j, i)$ wherein $x \in \Real^n$ is a state, $j \in [m]$ is a dynamic that has been assigned to the state $x$ by our tree search and $i \in [k]$ asserts that  the $i^{th}$ piece must be maximal at point $x$.   

\begin{defn}[Constraints for Set of Witnesses]
The set of witnesses for each node $W(\alpha)$ poses the following polyhedral constraint $CW(\alpha)$:
  \begin{equation}\label{eq:cw-alpha-constraint}
   \bigwedge_{(x, j, i) \in W(\alpha)} \left(\begin{array}{l} 
\bigwedge_{i' \in [k]} c_i^t x \geq c_{i'}^t x  \ \land\ \\ 
c_i^t x - d_i \geq 0  \ \land\ \\ 
c_i^t (A_j x + b_j) \leq -\lambda (c_i^t x - d_i)  
\end{array}\right) 
\end{equation}
\end{defn}

The first line expresses that the $i^{th}$ piece is maximal at $x$, the second line enforces that the CBF is non-negative and 
the third enforces a  decrease condition through the dynamics $j \in [m]$ at point $x$.   Each node has the property that
$C(\alpha)\ = \ CW(\alpha)\ \land\ \psi_0$, where $\psi_0$ is the constraint from Eq.~\eqref{eq:cd-bounds}. 

\noindent\textbf{Initializing the Tree:} The initial tree just has a root node $\alpha_0$ with $W(\alpha_0) = \emptyset$ and
$C(\alpha_0) = \psi_0$. 

The algorithm proceeds iteratively, with each iteration number $p$ consisting of three steps: (a) selecting a
candidate node (line~\ref{nl:choose-candidate}); (b) verifying the candidate (line~\ref{nl:verify-candidate}); and (c) refining the tree if verification fails (lines~\ref{nl:create-children} -~\ref{nl:define-c-beta}). 

\noindent\textbf{Selecting a Candidate:} At the beginning of the $p^{th}$ iteration, 
we select a candidate CBF $B^{(p)} = \max_{i \in [k]} c_i^{(p)} x - d_i^{(p)}$. This is performed by 
selecting a previously unexplored leaf $\beta$ of the tree. The leaf is then marked as explored. 
We then choose a point $(c_i^{(p)}, d_i^{(p)})_{i \in [k]}$ that satisfies $C(\beta)$. 
If $C(\beta)$ is infeasible, the algorithm simply moves on to itertaion $p+1$.

\noindent\textbf{Verifier:} The verifier receives a function $B^{(p)}$ and either needs to 
certify that $B^{(p)}$ is a CBF or find a counter-example $x \in \Real^n$ showing that
$B^{(p)}$ is not a CBF. To do so, we check conditions \textbf{(C1)}-\textbf{(C3)} ( Def.~\ref{def:polyhedral-cbf} ).

Recall that condition \textbf{(C1)} requires $B^{(p)}(x) \leq -\epsilon$ for all $x \in X_0$, wherein $\epsilon > 0$ is a user-provided
tolerance. Equivalently, we may write this as: $ \forall x \in X_0,\ \bigwedge_{i \in [k]} c_i^{(p)} x - d^{(p)}_i \leq - \epsilon$.
We solve a series of $k$ linear programs: 
$ \max\ c_i^{(p)} x - d^{(p)}_i + \epsilon \ \mathsf{s.t.}\ x \in X_0 $.
Let's assume that $X_0$ is non-empty and compact polyhedron. An optimal solution is guaranteed to exist.
If the optimal objective value for all the LPs are $\leq 0$,  we conclude that condition 
\textbf{(C1)} holds. Otherwise, we have a counterexample, $x_p$ that certifies that for some
$i \in [k]$, $c_i^{(p)} x_p - d^{(p)}_i  > -\epsilon$, thus violating \textbf{(C1)}.

Condition \textbf{(C2)} requires that $B^{(p)}(x) \geq \epsilon$ for all $x \in X_u$. Alternatively, we assert
that the set $C_0^{(p)} := \{ x \ |\ B^{(p)}(x) \leq 0 \}$ does not intersect the set $X_u$. Assuming that $X_u$ is a non-empty 
and compact polyhedral set, there must exist a separating hyperplane $c_p^t x - d_p $ such that 
$C_0^{(p)} \subseteq \{ x \ |\ c_p^t x - d_p \leq 0 \} $ and $ X_i \subseteq \{ x \ |\ c_p^t x - d_p \geq \epsilon \}$. 
A simply trick that saves computational time is to assert that the ``first piece'' $c_1^{(p)} x - d_1^{(p)}$ form this 
separating hyper plane. This is a sufficient condition that guarantees \textbf{(C2)}.

\begin{lem}\label{lem:c2-suff-cond}
Any candidate solution  $(c_i, d_i)_{i=1}^k$ that satisfies  condition \textbf{(C2p)} : 
$\forall\ x \in X_u,\ c_1^t x - d_1 \geq \epsilon$, 
satisfies \textbf{(C2)} from Def.~\ref{def:polyhedral-cbf}.
\end{lem}

Thus, to verify \textbf{(C2p)}, we solve the LP:
$\min (c_1^t x - d_1 - \epsilon)\ \mathsf{s.t.}\ x \in X_u$.
Assuming compactness and non-emptiness of $X_u$, we are guaranteed that the LP 
has an optimal solution. If this solution has objective value $< 0$, we conclude 
that \textbf{(C2p)} is violated with witness $x_p$ corresponding to the
optimal solution. Otherwise, we note that \textbf{(C2)} must hold by Lemma~\ref{lem:c2-suff-cond}.

Finally, we wish to verify condition \textbf{(C3)}. Recall that \textbf{(C3)} requires 
for every state $x$ such that $B(x) \leq 0$, there must exist a dynamical mode $l \in [m]$ such that every piece of the barrier function that is maximized at $x$ must satisfy a ``decrease condition''
We search for a state $x$ that violates \textbf{(C3)}: 
\begin{equation} \label{eq:c3-falsification}
\left. \begin{array}{l}
    \textsf{find}\   x \in \Real^n\ \mathsf{s.t.}\\
  \BigAnd_{l=1}^m \BigOr_{i=1}^k \ 
    \left( \begin{array}{l} (c^{(p)}_i)^t x - d^{(p)}_i \leq 0 \ \land\ \\ 
    \BigAnd_{i' \in [k]} (c^{(p)}_i)^t x - d^{(p)}_i \geq (c^{(p)}_{i'})^t x - d_{i'}^{(p)} \\
    (c_i^{(p)})^t (A_l x + b_l) \geq -\lambda ((c^{(p)}_i)^t x - d_i^{(p)}) \end{array} \right)
\end{array}\right\}
\end{equation}
Note that the constraint can be encoded as a mixed-integer optimization problem due to the presence of the disjunction (Appendix~\ref{app:milp=encoding}). If the MILP is infeasible, the condition \textbf{(C3)} is verified and
$B^{(p)}$ is the barrier function we seek. Otherwise, we obtain a witness $x_p$.

\noindent\textbf{Refining the Tree:} Suppose a leaf node $\beta$ were chosen as the node to be explored by the tree search, and the corresponding candidate 
$B^{(p)}$ resulted in a witness $x_p \in \Real^n$ that fails one of the verification conditions \textbf{(C1)}, \textbf{(C2p)} or 
\textbf{(C3)}, we will refine $\beta$ by adding children  $\beta_{i, j}$ for each $i \in [k]$ and $j \in [m]$. The child $\beta_{i,j}$ has the following 
attributes:
\begin{equation} \label{eq:cw-definition-child}
\begin{array}{l}
W(\beta_{i,j}):= W(\beta) \cup \{ ( x_p, j, i ) \},\ \text{and}\\
C(\beta_{i,j}) := C(\beta) \ \land\ CW( \{ (x_p, j, i ) \} ) \,.
\end{array}
\end{equation}

The constraint $C(\beta_{i,j})$ eliminates the candidate chosen at the parent node $\beta$: 
\begin{lem}\label{thm:candidate-exclusion-childen}
Let $(c_i^{(p)}, d_i^{(p)})_{i \in[k]}$ be the candidate with node $\beta$ at the $p^{th}$ iteration and $\beta_{i,j}$ be a child of $\beta$. then
$(c_i^{(p)}, d_i^{(p)} )_{i \in [k]} \not \in C(\beta_{i,j})$.
\end{lem}
\begin{pf}
Proof follows directly from the following three observations: $x_p$ is a counterexample to one of the 
CBF conditions for the candidate $(c_i^{(p)}, d_i^{(p)})_{i \in[k]}$. However, $CW(\beta)$ includes 
$ CW( \{ (x_p, j, i ) \} )$, which constraints all the solutions to $C(\beta_{i,j})$ of the child node $\beta_{i,j}$
to satisfy the CBF conditions at point $x_p$ with dynamic $j$ and piece $i$ being maximal. As a result, 
$(c_i^{(p)}, d_i^{(p)} )_{i \in [k]} \not \in C(\beta_{i,j})$.
\end{pf}
\begin{thm}
If Alg.~\ref{alg:tree-search-explore} returns a CBF, then it must satisfy constraints \textbf{(C1)-(C3)}.
\end{thm}
\begin{pf}
Proof follows by a direct inspection of the algorithm and the verification procedure. Note that line~\ref{nl:return-cbf} is the only place
that returns a CBF and it is executed only if the verification of \textbf{(C1)}-\textbf{(C3)} succeed.
\end{pf}
\noindent\textbf{Termination:} We adapt the argument in our previous work focusing on CLF synthesis to enable termination~\cite{berger2022learning}, as follows:
\begin{compactenum}
    \item The candidate $(c_i^{(p)}, d_i^{(p)})_{i \in [k]}$ is the center of the maximum volume inscribed ellipsoid (MVE-center) of the polyhedron $C(\beta)$ (line~\ref{nl:choose-candidate}). This ensures that for each child $\beta_{i, j}$, the volume of $C(\beta_{i,j}) \leq \gamma C(\beta)$ where $\gamma < 1$ is a volume shrinking factor. 
    \item We will stop exploration of a leaf node $\beta$ whenever $C(\beta)$ is empty or the Chebyshev radius of $C(\beta)$ is below a cutoff threshold  $R_{\min}$ (line~\ref{nl:termination-criterion}). 
\end{compactenum}

\begin{thm}
Algorithm~\ref{alg:tree-search-explore} terminates in $O\left( (nk)^{O( (n+1)^2 k^2)}\right)$ iterations, wherein 
each iteration's cost is dominated by solving a mixed-integer optimization problem corresponding to ~\eqref{eq:c3-falsification} with 
$O(mk)$ binary variables, $(n+1)k$ real-valued variables and $O(mk^2)$ constraints.
\end{thm} 

The full proof is provided in Appendix~\ref{app:termination-analysis}. 

\noindent\textbf{Synthesizing Min-Barrier Certificate:} The process of synthesizing barrier certificates of the form $\min_{i \in [k]}(c_i^t x - d_i)$ is very similar to that shown in Algorithm~\ref{alg:tree-search-explore}. The key differences include (a) the verification procedure needs to be modified appropriately to check the min-barrier condition in Section~\ref{sec:min-barriers}; and (b) the definition of $CW(W(\alpha))$ for a node $\alpha$ must be modified appropriately. We omit details due to lack of space.

\section{Min-Max Multiple Barriers}\label{sec:Multiple-barriers}
The exponential cost for Algorithm~\ref{alg:tree-search-explore} is not surprising given 
our previous work that has established the NP-hardness of elementary problems such as checking  polyhedral Lyapunov functions~\cite{berger2022learning}. However, 
compared to Lyapunov functions, the synthesis of CBFs has two distinct advantages: (a) CBFs establish control invariance for a subset of the state-space: as a result, 
we can attempt to synthesize CBFs with small values of $k$ such as $k =2$, with the caveat that they may yield relatively small control invariant sets; and (b) 
to offset the disadvantage, we can combine the control invariant region $ CI(B_j) :=  \{ x\ |\ B_j(x) \leq 0 \}$ for multiple control barrier functions. 
Let $B_{j}$ for $j = 1,\ldots ,N$ be a set of CBFs synthesized by our algorithm. These could include barriers formed by $\max$ and $\min$ of affine functions.
Note that $\mathcal{C} = \bigcup_{j=1}^N CI(B_j)$ is a control invariant region, since for any $x \in CI(B_j)$, we have a control strategy that keeps our system inside 
$CI(B_j)$. The region $\mathcal{C}$ itself can be viewed as the zero sublevel set of the function $B(x) = \min_{j=1}^N(B_j(x))$. This yields a so-called ``mini-max''
control barrier function obtained by combining $\min$ and $\max$ barrier functions.

\noindent\textbf{Computing Multiple CBFs:} In order to compute multiple CBFs, we adapt the following strategy proposed by Wajid and Sankaranarayanan for the case of polynomial CBFs~\cite{Wajid+Sankaranarayanan/2025/Successive}. As inputs to the CBF problem, we have an unsafe set $X_u$ but do not include an initial set $X_0$ since our goal  is to 
compute as large a control invariant set as possible. To do so, we select a set of random test ponts $X_t = \{ x_1, \ldots, x_N \}$ wherein $x_i \not\in X_u$. Next, 
we apply Algorithm~\ref{alg:tree-search-explore} or its counterpart for $\min-$barrier functions using $X_0  = \{ x_i \} $ in order to attempt synthesis of a CBF
that \emph{separates} $x_i$ from the unsafe set $X_u$. The final result is a union of all the control invariant sets thus obtained.

\section{Empirical Results}\label{sec:Examples}
In this section, we use our algorithm to synthesize a min-max control barrier function for some switched systems. At first, we study a numerical example. Then we consider a DC-DC converter as a switched affine system to synthesize control barrier functions and switching controllers. Finally, we will utilize the algorithm for higher order systems as a multi-agent system and car's velocity which are modeled by switched dynamics.  Table~\ref{tab:empirical-results} shows the results from these benchmarks at a glance.

\begin{table*}[h!]
\caption{Summary of results for the empirical evaluation. $\max$ refers to $\max$-CBFs while $\min$ refers to $\min$-CBFs. $D_{\max}$ refers to cutoff depth for  the tree search.}\label{tab:empirical-results}
{\footnotesize 
\begin{center}
\begin{tabular}{|c|c|c|c|c|c|c|c|c|c|c|}
\hline
\textbf{ID} & $n$ & $m$ & \multicolumn{2}{c}{\#\textbf{Test Pts.}} \vline & \multicolumn{2}{c}{\#\textbf{CBFs}} \vline & \multicolumn{2}{c}{$D_{\max}$} \vline & 
\multicolumn{2}{c}{\textbf{ Time (sec)}} \vline\\ 
\cline{4-11}
& & & $\max$ & $\min$ & ${\max}$ & $\min$ & ${\max}$ & ${\min}$ & ${\max}$ & ${\min}$ \\\hline
Ex.\ref{example:numerical} & 2 & 3 & 11 & 11 & 11 & 11 & 6 & 6 & 53.5 & 20.4 \\\hline
Ex.\ref{Example:DCDC} & 2 & 2 & 16 & 14 & 16 & 14 & 6 & 6 & 16.3 & 6.8 \\\hline
Ex.\ref{Ex:Multi-agent} & 3 & 9 & 6 & 6 & 6 & 6 & 6 & 6 & 26.6 & 31.8 \\\hline
Ex.\ref{Ex:Car} & 6 & 6 & 7 & - & 7 & - &  14 & -  & 885.61 & -\\
\hline
\end{tabular}
\end{center}
}
\end{table*}

\begin{exmp} \label{example:numerical}  
    Consider a system with $2$  state variables and $3$ dynamical modes to choose from:
    \begin{equation}
    \nonumber
        A_1 = \begin{bmatrix}
            1 & -1 \\ -0.5 & -2
        \end{bmatrix}, ~ 
        A_2 = \begin{bmatrix}
            -2 & 1 \\ 
            0.5 & -1
        \end{bmatrix}, ~
        A_3 = \begin{bmatrix}
            1 & 1 \\ -1 & -1
        \end{bmatrix}
    \end{equation}
    \begin{equation}
    \nonumber
        b_1 = \begin{bmatrix}
            -1 \\ -1
        \end{bmatrix}, ~ 
        b_2 = \begin{bmatrix}
            1 \\ -1
        \end{bmatrix}, ~
        b_3 = \begin{bmatrix}
            1 \\ 1
        \end{bmatrix}
    \end{equation}
   The unsafe set is defined as $X_u = [-0.5, 0.5] \times [-0.5, 0.5]$. Using the proposed algorithm, we synthesize a min-max multiple barrier function with $k = 2$. We consider various test points $x_t \not\in X_u$ a and for each of them we attempt to find $c_i$ and $d_i$ for a CBF that ``separates'' $x_t$ as the initial point from the unsafe set $X_u$. Tables \ref{tab:numerical-example1} and table \ref{tab:numerical-example2} in Appendix~\ref{app:numerical-tables} reports the $\min$ and $\max$ CBFs obtained by running Algorithm~\ref{alg:tree-search-explore}. For efficiency, we truncate the search whenever the depth of the tree exceeds $D_{\max} = 6$. We synthesized $11$ max barrier functions and $11$ min barrier functions, in all.  
    Figure~\ref{fig:result-Numerical-exm} shows the synthesized min-max barrier functions and simulations of the trajectories using the corresponding switching rule for randomly chosen initial conditions. 

\begin{figure}
    \centering
    \includegraphics[width=1\linewidth]{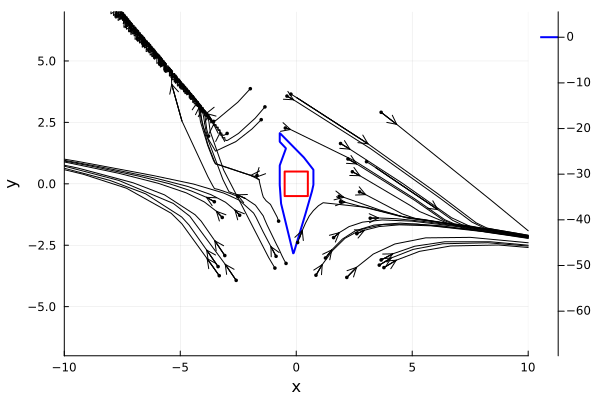}
    \caption{Zero level set of the min-max barrier $B^*(x) = 0$ (blue lines), unsafe set (red lines), the trajectories for each initial conditions (black lines)}
    \label{fig:result-Numerical-exm}

\end{figure}
\end{exmp}

\begin{exmp}[DC-DC Converter]
\label{Example:DCDC} The dynamics for this DC-DC converter are taken from Ravanbakhsh et al~\cite{ravanbakhsh2017class}. The switched model of this system has two modes ($l =2$) with dynamics $\dot{x}(t) = A_l x(t) + b_l$ in which $x(t)$ denotes the state variables $[i(t), v(t)]$. $i(t)$ and $v(t)$ are current and voltage of the converter respectively. 
    \begin{equation}
        \nonumber
        l = 1: ~~ A_1 = \begin{bmatrix}
            0.0167 & 0 \\ 0 & -0.0142 
        \end{bmatrix}, ~~~ b_1 = \begin{bmatrix}
            0.3333 \\ 0
        \end{bmatrix}
    \end{equation}
    \begin{equation}
        \nonumber
        l = 2: ~~ A_2 = \begin{bmatrix}
            -0.0183 & -0.0663 \\ -0.0711 & -0.0142
        \end{bmatrix}, ~~~ b_2 = \begin{bmatrix}
            0.3333 \\ 0
        \end{bmatrix}
    \end{equation}
    We assume the unsafe set $X_u = [0,1] \times [0,1]$. By considering the number of the pieces of the polyhedral function $k=2$, and applying the proposed algorithm with a depth cutoff of $6$ for $16$ test points which yielded $14$ CBFs each for the $\max$ and $\min$ cases. The coefficients $c_i$ and $d_i$ of the polyhedral barriers for the minimum and maximum functions are shown in Tables~\ref{tab:DCDC-example-max} and ~\ref{tab:DCDC-example-min} (Appendix~\ref{app:numerical-tables}), respectively. The final barrier depicted in Figure~\ref{fig:DC-DC-Converter} combines all the barriers synthesized.

    \begin{figure}
        \centering
        \includegraphics[width=1\linewidth]{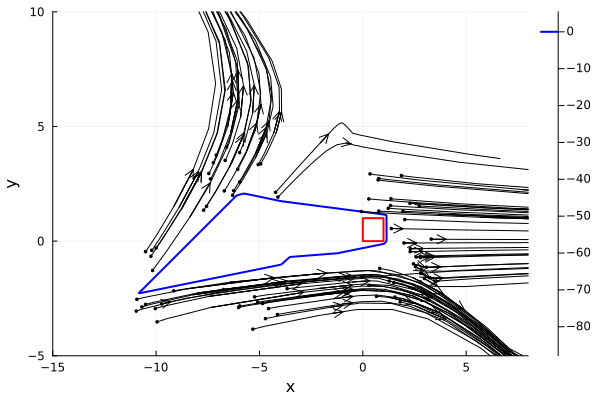}
        \caption{Zero level set of the min-max barrier $B^*(x) = 0$ (blue lines), unsafe set (red lines), trajectories for various initial conditions (black lines)}
        \label{fig:DC-DC-Converter}
    \end{figure}

    \end{exmp}
    
\begin{exmp}[Multi-agent System] Consider a multi-agent system with the following model: 
\label{Ex:Multi-agent}
\begin{equation}
\nonumber
    \dot{x}(t) = \Sigma_{i=1}^n u_i(t) g_i(x(t)). 
    \label{eq:multi}
\end{equation}
We use a 3-dim Brockett integrator with two inputs \cite{Brockett1982}:
\begin{equation}
\begin{cases}
\nonumber
    \dot{x}_1(t) = u_1(t),\      \dot{x}_2(t) = u_2(t)\\
    \dot{x}_3(t) = x_1 u_2 - x_2 u_1.
    \label{eq3dim}
\end{cases}
\end{equation}
In which, $x_i(t) \in R$ is a state that shows the location of $i^{th}$ agent, and $u_i(t) \in R$ is a control signal applied to the corresponding agent. It has been proved that this non holonomic system cannot be asymptotically stabilized by a continuous feedback law \cite{liberzon2003switching}. So, we are considering a piece-wise constant control signal $u_1(t) \in \{ -\alpha, 0 , \alpha \}$ and $u_2(t) \in \{ -\beta, 0, \beta\}$ with constants $\alpha, \beta \in \mathbb{R}_{>0}$. By this control signals, the system is modeled as a switched affine system $\dot{x}(t) = A_{i,j} x(t) + b_{i,j}$, wherein $i \in [3], j \in [3]$ reflect the fixed choices of $u_1, u_2$. 
In this example we consider $\alpha = 1$ and $\beta = 1$ and the unsafe set $[0,1] \times [0,1] \times [0,1]$. The table \ref{tab:multi-agent} (Appendix~\ref{app:numerical-tables}) shows the CBFs discovered for each test point with $k = 3$ and maximum depth cutoff of 6.
Table~\ref{tab:empirical-results} summarizes the performance of our approach on this benchmark. 

\end{exmp}

\begin{exmp}[Vehicle Model]
\label{Ex:Car}
For this example, consider the follwing dynamic. 
\begin{equation*}
      \left\{ \begin{array}{cc} 
      \begin{array}{l}
            \dot{x} = V_x,\\
            \dot{y} = V_y,\\
            \dot{z} = V_z,\\
            \end{array} & 
            \begin{bmatrix}
                \dot{V}_x \\
                \dot{V}_y \\
                \dot{V}_z
            \end{bmatrix} = A_{3 \times 3} \begin{bmatrix}
                V_x - V_{xref} \\
                V_y - V_{yref} \\
                V_z - V_{zref}
            \end{bmatrix} 
    \end{array}\right.
    \end{equation*}
wherein 
\begin{equation*}
    A = \begin{bmatrix}
        -2.5 & 0.8 & 1.2\\ -1.5 & -2.2 & 0.7 \\ 0.5 & -1.8 & -2.7
    \end{bmatrix}.
\end{equation*}
To control the movement, the references' velocities can apply in various directions which results in a switched affine systems. In this example, we have considered 6 directions by defining $[V_{xref}; V_{yref}; V_{zref}]$ as
\begin{equation*}
    \begin{bmatrix}
        1 \\ 0 \\ 0
    \end{bmatrix}, ~
    \begin{bmatrix}
        -1 \\ 0 \\ 0       
    \end{bmatrix},~
    \begin{bmatrix}
        0 \\ 1 \\ 0
    \end{bmatrix},~
    \begin{bmatrix}
        0 \\ -1 \\ 0
    \end{bmatrix},~
    \begin{bmatrix}
        0 \\ 0 \\ 1
    \end{bmatrix},~
    \begin{bmatrix}
        0 \\ 0 \\ -1
    \end{bmatrix}
\end{equation*}
which results in a switched affine systems with 6 subsystems. The given unsafe set is $[-0.5, 0.5]^3 \times [-10, 10]^3$. Some of the found barrier functions for some test points have been shown in the table \ref{tab:Ex-Car} (Appendix~\ref{app:numerical-tables}).

\end{exmp}

\section{Conclusion}  

To conclude, we have demonstrated an approach that defines a class of  non-smooth control barrier functions for switched affine systems through 
the application of piecewise $\max$ and $\min$ over affine functions. We have provided algorithms for synthesizing these control barrier function 
and demonstrated a scheme for extracting a feedback control from a CBF. Our future work seeks to expand our work to polynomial dynamics using 
ideas such as Sum-of-Squares (SOS) programming. We are also interested in the use of piecewise linearization of nonlinear dynamics, especially 
for applications to robust plan execution in robotic systems.

\begin{ack}                               
Will be provided in subsequent version. 
\end{ack}

\bibliographystyle{plain}       
\bibliography{autoReferences}

\appendix
\section{Encoding Verification as a Mixed-Integer Linear Program}\label{app:milp=encoding}
We will now encode the constraint for finding a potential counterexample for condition \textbf{(C3)} as a mixed-integer optimization problem. This constraint corresponds to 
Eq.~\eqref{eq:c3-falsification} and is recalled below:
\begin{equation} \label{eq:c3-fals-app}
\left. \begin{array}{l}
    \textsf{find}\   x \in \Real^n\ \mathsf{s.t.}\\
  \BigAnd_{l=1}^m \BigOr_{i=1}^k \ 
    \left( \begin{array}{l} (c^{(p)}_i)^t x - d^{(p)}_i \leq 0 \ \land\ \\ 
    \BigAnd_{i' \in [k]} (c^{(p)}_i)^t x - d^{(p)}_i \geq (c^{(p)}_{i'})^t x - d_{i'}^{(p)} \\
    (c_i^{(p)})^t (A_l x + b_l) \geq -\lambda ((c^{(p)}_i)^t x - d_i^{(p)}) \end{array} \right)
\end{array}\right\}
\end{equation}

To encode this as a MILP, we first introduce binary variables of the form 
$w_{l, i} \in \{ 0, 1\}$ for $ l = 1, \ldots, m$ and $i = 1, \ldots, k$.
If $w_{l,i} = 1$ for some specific value of $l, i$, we will ensure that the inequalities shown above will hold for that
$l, i$. In other words, we wish to satisfy:

\[ \left( \begin{array}{l} 
w_{l, i} = 1\ \Rightarrow\ (c^{(p)}_i)^t x - d^{(p)}_i \leq 0 \ \land\ \\ 
    \BigAnd_{i' \in [k]} w_{l, i} = 1\ \Rightarrow\ (c^{(p)}_i)^t x - d^{(p)}_i \geq (c^{(p)}_{i'})^t x - d_{i'}^{(p)} \\
    w_{l,i} = 1\ \Rightarrow\ (c_i^{(p)})^t (A_l x + b_l) \geq -\lambda ((c^{(p)}_i)^t x - d_i^{(p)}) \end{array}\right) \]

Additionally, we enforce the conjunction of disjunctions by the constraint:
\[ w_{l,1} + \cdots + w_{l, k} \geq  1,\ \mathbf{for}\ l = 1, \ldots, m\]

The implication is encoded using a $M$ constraint. Consider a constraint of the form 
\[ w_{l,i} = 1 \ \Rightarrow\  c^t x - d \geq 0 \ ,.\] 
We encode this by introducing a constraint of the form: 
\[ c^tx - d \geq -(M \|c\|_1 + \| d\|_1 ) (1-w_{l,i}) \,. \]
Here $M$ is chosen so that the following property holds:
If there exists a feasible solution $x$ to ~\eqref{eq:c3-fals-app} then there exists a solution with  $\| x \|_{\infty} \leq M$. 
Note that if $w_{l,i}$ equals $1$, we obtain the constraint $c^t x - d \geq 0$. Otherwise, we obtain the constraint
$c^t x - d \geq - (M \|c\|_1 + \|d\|_1)$, which is simply a consequence of $\|x \|_{\infty} \leq M$.

Since, the candidate $(c^{(p)}_i, d^{(p)}_i)_{i=1}^k$ is fixed. Let $a = \max_{i=1}^k \|c^{(p)}_i \|_1$.  A bound $M$ over the magnitude of $\|x_c\|_{\infty}$ is obtained as follows:

\begin{lem}\label{lem:counterex-big-M-bound}
Let $x_c \in \Real^n$ be a solution to~\eqref{eq:c3-fals-app}. Let $a = \max_{i=1}^k \max( \|c^{(p)}_i\|_1, d^{(p)}_i) $ and $U = \max(2, {\lambda} + \|A_1 \|_{\infty}, \ldots, {\lambda} + \|A_m\|_{\infty} ) a $. It follows that $\|x_c\|_{\infty} \leq (n U)^n$.
\end{lem}
\begin{pf}
Note that by~\eqref{eq:c3-fals-app}, there exists a map $\mu: [m] \rightarrow [k]$ and a number $\epsilon > 0 $ such that $x_c$ belongs to a polyhedron defined by inequalities of the form:
\begin{equation}\label{eq:counterex-polyhedral-form} 
\bigwedge_{l=1}^m \left( \begin{array}{c}
c^{(p)}_{\mu(l)} x_c \leq d^{(p)}_{\mu(l)} \\ 
\bigwedge_{j=1}^k (c^{(p)}_{\mu(l)} - c^{(p)}_j)^t x_c \geq d^{(p)}_{\mu(l)} - d^{(p)}_j \\ 
(c^{(p)}_{\mu(l)})^t (A_l + \lambda I) x_c \geq \epsilon - (c^{(p)}_{\mu(l)})^t b_l + \lambda d^{(p)}_{\mu(l)}
\end{array}\right) 
\end{equation}
A well known result in linear programming (Cf. Bertsimas and Tsitsiklis, Ch. 8~\cite{Bertsimas+Tsitsiklis/1997/Introduction}) states that for any polyhedron $ A x \leq b$, where $x$ is a $n \times 1$ vector of unknowns, if there is a feasible point then there exists a feasible point that satisfies $ \| x \|_{\infty} \leq (n U)^n$,
wherein $U$ is the absolute value of the largest entry in $A,b$. The result follows by careful examination of the constraints from~\eqref{eq:counterex-polyhedral-form}. Note that $\|c_i\|_{\infty} \leq \gamma$ and $\|(c^{(p)}_j)^t (A_l + \lambda I) \|_\infty \leq \|c^{(p)}_j\|_1 ( |\lambda| + \|A_l\|_{\infty})$.
\end{pf}

\section{Termination and Complexity Analysis of Algorithm~\ref{alg:tree-search-explore}} \label{app:termination-analysis}
In this section, we will present an analysis of Algorithm~\ref{alg:tree-search-explore} to prove its termination, based on setting parameters
$\gamma > 0$ that sets the initial search region for the coefficients of the candidate CBF (see Eq.~\eqref{eq:cd-bounds}) and the termination criterion 
that cuts off search whenever the Chebyshev radius of node $\alpha$ is below $R_{\min}$. 

Let $\alpha_0$ be the root of our search tree. Its volume $V_0 = (2 \gamma)^{k (n+1)}$. Next, let $\alpha$ be any node explored by the algorithm. 
Since we require the Chebyshev radius of $CW(\alpha)$ to be at least $R_{min}$, we have the following lemma.
\begin{lem}
If a polyhedron $P$ has Chebyshev radius at least $R_{\min}$ then its volume $\mathsf{vol}(P) \geq (K R_{\min})^{k(n+1)}$ for some constant $K$.
 \end{lem}
 \begin{pf}
Proof follows from the observation that $P$ must contain a sphere of radius at least $R_{\min}$ over $k(n+1)$ dimensions.
 \end{pf}

 Next, consider two nodes in the tree, parent $\alpha$ and its immediate child $\beta$.  We have the following relationship between the volumes of the 
 polyhedra $CW(\alpha)$ and $CW(\beta)$.

\begin{lem}\label{lem:mve-center-exclusion}
Let $(c_i, d_i)_{i=1}^k$ represent the MVE center of $CW(\alpha)$. It follows that $CW(\beta) \subseteq CW(\alpha)$ and $(c_i, d_i)_{i\in [k]} \not\in CW(\beta)$.
 \end{lem}
 \begin{pf}
Since $\beta$ is the child of $\alpha$, we note that $CW(\alpha)$ is feasible (see Algorithm~\ref{alg:tree-search-explore}, line~\ref{nl:termination-criterion}).
The MVE center $(c_i, d_i)_{i=1}^k$ of $CW(\alpha)$ is chosen as the candidate. Also, since we created $\beta$, this candidate must have failed verification.
Therefore, following Theorem~\ref{thm:candidate-exclusion-childen}, we obtain that  $(c_i, d_i)_{i=1}^k \not\in CW(\beta)$. Also, by definition of 
$CW(\beta)$ (Eq.~\eqref{eq:cw-definition-child}, we note that $CW(\beta) \subseteq CW(\alpha)$.
 \end{pf}

 \begin{lem}
The volumes of the polyhedra $CW(\alpha)$ and $CW(\beta)$ satisfy the relationship
\[ \mathsf{vol}(CW(\beta)) \leq (1 - \gamma) \mathsf{vol}(CW(\alpha)) \,,\]
wherein $\gamma = \frac{1}{ (n+1)k}$.
 \end{lem}
 \begin{pf}
Proof follows from the so-called ``cutting-plane'' argument first proposed by Tarasov et al and explained by Boyd and Vandenberghe.
Since $CW(\beta) \subseteq CW(\alpha)$ and excludes its MVE-center, we have 
\[ \mathsf{vol}(CW(\beta)) \leq (1 - \gamma) \mathsf{vol}(CW(\alpha)) \,,\]
wherein $\gamma = \frac{1}{ (n+1)k}$.
 \end{pf}

\begin{lem}
The depth of any branch of the search tree cannot exceed the bound: 
$O((n+1)^2 k^2)$.
\end{lem}
\begin{pf}
Note that for each branch the volume of the polyhedron at the root is $V_0: (2\gamma)^{nk+k}$  and the search terminates whenever 
a descendant $\alpha$ has $\mathsf{vol}(CW(\alpha)) \leq K R_{\min}^{nk + k}$. Furthermore, at each step, the volume of a child node
is at most $1 - \gamma$ times the volume of its parent. Combining, we obtain that the volume of a node  $\alpha_D$ at depth $D$ must be bounded by
\[ K R_{\min}^{nk + k} \leq \mathsf{vol}(\alpha_D) \leq (1-\gamma)^D V_0 \,.\]
Therefore, we have 
\begin{align*} 
D & \leq \frac{\log (KR_{\min})^{nk +k} - \log (2\gamma)^{nk+k}}{\log(1-\gamma)} \\
& \leq  \frac{1}{\gamma} (nk+k) (\log ( K R_{\min}) - \log (2\gamma) )  =O( (n+1)^2 k^2)
\end{align*}
Note that we employ the inequality $ -\log(1-\frac{1}{r}) \geq \frac{1}{r}$ for $r \geq 2$.
\end{pf}

\begin{thm}
Algorithm~\ref{alg:tree-search-explore} terminates in $O\left( (nk)^{O( (n+1)^2 k^2)}\right)$ iterations, wherein 
each iteration's cost is dominated by solving a mixed-integer optimization problem with 
$O(mk)$ binary variables, $(n+1)k$ real-valued variables and $O(mk^2)$ constraints.
\end{thm} 
\begin{pf}
Proof is by  observing that the search tree has depth at most $D = O((n+1)^2 k^2)$ and branching factor
of at most $nk$. The cost at each step is dominated by solving a verification problem which involves solving a MILP.

\end{pf}
\section{Details of CBFs for Numerical Examples}\label{app:numerical-tables}
The coefficients for various CBFs from Example~\ref{example:numerical}, \ref{Example:DCDC}, \ref{Ex:Multi-agent}, and \ref{Ex:Car} are reported in the following tables.

\begin{table}
    \centering
    \caption{The coefficients of the minimum functions corresponding to each test point for the example \ref{example:numerical}}
    \label{tab:numerical-example2}
    \begin{tabular}{|c|c|}
        \hline
        \textbf{$x_t$} & \textbf{$[c_i, d_i]$} \\
        \hline
        [1; 0] & \makecell{$i=1: [-7.32068;\; 0.0;\; -5.43091]$ \\ $i=2: [-9.37083;\; 0.88549;\; -9.37083]$} \\
        \hline
        [2; 4] & \makecell{$i=1: [-4.02462;\; -2.95583;\; -7.52341]$ \\ $i=2: [-5.22319;\; 1.75726;\; -7.52341]$} \\
        \hline
        [-1; 0] & \makecell{$i=1: [9.04427;\; 0.0;\; -6.69266]$ \\ $i=2: [-9.04427;\; 1.19060;\; -9.04427]$} \\
        \hline
        [-1; -1] & \makecell{$i=1: [8.83221;\; 2.07361;\; -7.88315]$ \\ $i=2: [-0.83174;\; -0.64777;\; -8.83221]$} \\
        \hline
        [0; -3] & \makecell{$i=1: [-9.79703;\; 3.11558;\; -7.70489]$ \\ $i=2: [-5.85562;\; 2.26523;\; -9.79703]$} \\
        \hline
        [1; -1] & \makecell{$i=1: [-8.97656;\; 1.07566;\; -7.27957]$ \\ $i=2: [-8.97656;\; -0.63089;\; -8.97656]$} \\
        \hline
        [1; 1] & \makecell{$i=1: [-8.64905;\; -3.33991;\; -8.64905]$ \\ $i=2: [1.41747;\; -0.34003;\; -8.64905]$} \\
        \hline
        [-1; 1] & \makecell{$i=1: [9.44861;\; 0.72932;\; -6.76427]$ \\ $i=2: [-0.55924;\; 0.01527;\; -9.44861]$} \\
        \hline
        [0; -4] & \makecell{$i=1: [-9.76948;\; 2.31929;\; -7.32671]$ \\ $i=2: [-9.76948;\; 0.55230;\; -9.76948]$} \\
        \hline
        [0.1; -5] & \makecell{$i=1: [-9.55313;\; 1.92169;\; -7.28471]$ \\ $i=2: [-4.34004;\; 1.26382;\; -4.34924]$} \\
        \hline
        [0; 2] & \makecell{$i=1: [-6.02489;\; -6.28970;\; -8.72223]$ \\ $i=2: [-1.11797;\; -0.52846;\; -8.72223]$} \\
        \hline
    \end{tabular}
\end{table}

\begin{table}
    \centering
    \caption{The coefficients of the maximum functions corresponding to each test point for the example \ref{example:numerical}}
    \label{tab:numerical-example1}
    \begin{tabular}{|c|c|}
        \hline
        \textbf{$x_t$} & \textbf{$[c_i, d_i]$} \\
        \hline
        [1; 0] & \makecell{$i=1: [-9.04427;\; 0.0;\; -6.69266]$ \\ $i=2: [-9.04427;\; 9.04427;\; 9.04427]$} \\
        \hline
        [2; 4] & \makecell{$i=1: [-8.17889;\; -0.08783;\; -7.36375]$ \\ $i=2: [-8.17889;\; 3.79783;\; 8.17889]$} \\
        \hline
        [-1; 0] & \makecell{$i=1: [9.04427;\; 0.0;\; -6.69266]$ \\ $i=2: [9.04427;\; 9.04427;\; 9.04427]$} \\
        \hline
        [-1; -1] & \makecell{$i=1: [9.22033;\; 0.0;\; -6.56506]$ \\ $i=2: [4.32697;\; -9.22033;\; 7.24379]$} \\
        \hline
        [0; -3] & \makecell{$i=1: [-9.86560;\; 3.00896;\; -7.60188]$ \\ $i=2: [-4.35311;\; -2.54928;\; 9.07284]$} \\
        \hline
        [1; -1] & \makecell{$i=1: [-8.43373;\; 3.55523;\; -8.64905]$ \\ $i=2: [-4.90758;\; -1.24332;\; -0.32435]$} \\
        \hline
        [1; 1] & \makecell{$i=1: [-8.64905;\; -3.33991;\; -8.64905]$ \\ $i=2: [-8.64905;\; -8.64905;\; 8.64905]$} \\
        \hline
        [-1; 1] & \makecell{$i=1: [8.64905;\; -3.33991;\; -8.64905]$ \\ $i=2: [8.64905;\; 8.64905;\; 3.64905]$} \\
        \hline
        [0; -4] & \makecell{$i=1: [9.83784;\; 2.53591;\; -8.47501]$ \\ $i=2: [9.83784;\; 9.83784;\; 9.83784]$} \\
        \hline
        [0.1; -5] & \makecell{$i=1: [-9.50073;\; 2.41938;\; -9.50073]$ \\ $i=2: [-4.09201;\; 1.47737;\; -4.24978]$} \\
        \hline
        [0; 2] & \makecell{$i=1: [-7.43252;\; -6.11803;\; -9.00001]$ \\ $i=2: [-0.18737;\; 2.28573;\; 7.80751]$} \\
        \hline
    \end{tabular}
\end{table}

\begin{table}[h!]
\centering
\caption{The coefficients of the polyhedral functions corresponding to each test point for DC-DC converter example (Maximum functions)}
\label{tab:DCDC-example-max}
\begin{tabular}{|c|c|}
    \hline
    \textbf{$x_t$} & \textbf{$[c_i, d_i]$ of maximum functions} \\
    \hline
    [2; 2] & \makecell{$i=1: [-6.73590;\; 0.29807;\; -9.06218]$ \\ $i=2: [-9.06218;\; -9.06218;\; 9.06218]$} \\
    \hline
    [2; -2] & \makecell{$i=1: [-1.61660;\; 6.94170;\; -6.94170]$ \\ $i=2: [-6.94170;\; 6.94170;\; 6.94170]$} \\
    \hline
    [-2; -2] & \makecell{$i=1: [-2.18600;\; 8.21615;\; -5.70875]$ \\ $i=2: [-3.21573;\; 2.28343;\; 3.21615]$} \\
    \hline
    [-2; 2] & \makecell{$i=1: [-0.44123;\; -6.64260;\; -9.29863]$ \\ $i=2: [-6.14976;\; -9.29863;\; -3.19363]$} \\
    \hline
    [0; 2] & \makecell{$i=1: [-0.45105;\; -6.06248;\; -9.08989]$ \\ $i=2: [-9.08989;\; 0.65610;\; 4.34728]$} \\
    \hline
    [2; 0] & \makecell{$i=1: [-6.21337;\; 0.84574;\; -8.84574]$ \\ $i=2: [-8.84574;\; 8.84574;\; 8.84574]$} \\
    \hline
    [0; -2] & \makecell{$i=1: [-0.57857;\; 2.02394;\; -2.14782]$ \\ $i=2: [-3.16742;\; -3.84871;\; 9.59748]$} \\
    \hline
    [-4; 3] & \makecell{$i=1: [-0.89956;\; -7.69256;\; -9.85070]$ \\ $i=2: [-1.84924;\; -9.85070;\; -9.85070]$} \\
    \hline
    [-6; -2] & \makecell{$i=1: [-1.92159;\; 9.53665;\; -3.57687]$ \\ $i=2: [-2.86645;\; 5.81445;\; 9.53665]$} \\
    \hline
    [-3; -1] & \makecell{$i=1: [-0.66285;\; 9.32860;\; -4.11326]$ \\ $i=2: [-3.51116;\; 4.43168;\; 9.32860]$} \\
    \hline
    [-10; 2] & \makecell{$i=1: [9.65532;\; -8.16787;\; -9.65532]$ \\ $i=2: [1.16079;\; -1.28966;\; -9.65532]$} \\
    \hline
    [-2; 3] & \makecell{$i=1: [-2.82248;\; -5.70603;\; -9.82742]$ \\ $i=2: [-9.82742;\; -5.69339;\; 4.22039]$} \\
    \hline
    [-1; 2] & \makecell{$i=1: [-1.22196;\; -6.46049;\; -9.51774]$ \\ $i=2: [-9.51774;\; -9.51774;\; -4.19785]$} \\
    \hline
    [-3; 2] & \makecell{$i=1: [-1.21686;\; -7.47827;\; -9.88841]$ \\ $i=2: [-2.64888;\; 0.26212;\; 9.88841]$} \\
    \hline
    [-5; -2] & \makecell{$i=1: [-2.04149;\; 9.35350;\; -3.95627]$ \\ $i=2: [-3.07436;\; 5.27966;\; 9.35350]$} \\
    \hline
    [-7; -2] & \makecell{$i=1: [-2.08143;\; 9.87401;\; -3.25863]$ \\ $i=2: [-3.04115;\; 6.66717;\; 9.87401]$} \\
    \hline
\end{tabular}
\end{table}

\begin{table}[h!]
\centering
\caption{The coefficients of the polyhedral functions corresponding to each test point for DC-DC converter example (Minimum functions)}
\label{tab:DCDC-example-min}
\begin{tabular}{|c|c|}
    \hline
    \textbf{$x_t$} & \textbf{$[c_i, d_i]$ of minimum functions} \\
    \hline
    [2; -2] & \makecell{$i=1: [-3.98740;\; 3.58889;\; -7.92371]$ \\ $i=2: [-1.67349;\; -0.23760;\; -7.92371]$} \\
    \hline
    [-2; -2] & \makecell{$i=1: [-5.64347;\; 9.83325;\; -6.87930]$ \\ $i=2: [-6.46475;\; -2.07967;\; -9.83325]$} \\
    \hline
    [-2; 2] & \makecell{$i=1: [-1.07170;\; -6.93520;\; -9.63650]$ \\ $i=2: [0.69027;\; -2.72209;\; -9.63650]$} \\
    \hline
    [0; 2] & \makecell{$i=1: [-0.67767;\; -6.02067;\; -9.15754]$ \\ $i=2: [-5.91852;\; -0.77982;\; -9.15754]$} \\
    \hline
    [2; 0] & \makecell{$i=1: [-6.21337;\; 0.84574;\; -8.84574]$ \\ $i=2: [-5.92143;\; 0.07492;\; -8.84574]$} \\
    \hline
    [0; -2] & \makecell{$i=1: [-8.74252;\; 5.56591;\; -9.89335]$ \\ $i=2: [-6.50007;\; -2.20855;\; -9.89335]$} \\
    \hline
    [-4; 3] & \makecell{$i=1: [-1.38943;\; -6.21646;\; -9.48972]$ \\ $i=2: [-1.68936;\; -5.91653;\; -9.48972]$} \\
    \hline
    [-6; -2] & \makecell{$i=1: [-2.33756;\; 9.83325;\; -3.57339]$ \\ $i=2: [-6.46475;\; -2.07967;\; -9.83325]$} \\
    \hline
    [-3; -1] & \makecell{$i=1: [-1.79642;\; 9.85791;\; -2.99737]$ \\ $i=2: [-8.57600;\; -0.03579;\; -9.85791]$} \\
    \hline
    [-2; 3] & \makecell{$i=1: [-0.58140;\; -5.14823;\; -8.80296]$ \\ $i=2: [-6.11009;\; 0.38046;\; -8.80296]$} \\
    \hline
    [-1; 2] & \makecell{$i=1: [-0.94552;\; -6.40921;\; -9.39779]$ \\ $i=2: [-7.54614;\; 0.19140;\; -9.39779]$} \\
    \hline
    [-3; 2] & \makecell{$i=1: [-1.01485;\; -7.34399;\; -9.76532]$ \\ $i=2: [-1.40860;\; -6.95024;\; -9.76532]$} \\
    \hline
    [-5; -2] & \makecell{$i=1: [-2.75289;\; 9.83325;\; -3.98871]$ \\ $i=2: [-6.46475;\; -2.07967;\; -9.83325]$} \\
    \hline
    [-7; -2] & \makecell{$i=1: [-2.02566;\; 9.83325;\; -3.26149]$ \\ $i=2: [-6.46475;\; -2.07967;\; -9.83325]$} \\
    \hline
\end{tabular}
\end{table}

\begin{table}[h!]
\centering
\caption{The coefficients of the polyhedral functions corresponding to each test point for the multi-agent system}
\label{tab:multi-agent}
\begin{tabular}{|c|c|}
    \hline
    \textbf{$x_t$} & \textbf{$[c_i, d_i]$ for the maximum functions} \\
    \hline
    [0; 4; 0] & \makecell{
        $i=1$: [9.54963;\; -3.10164;\; -1.33487;\; -9.54963] \\
        $i=2$: [-9.54963;\; -9.54963;\; 6.18004;\; 9.54963] \\
        $i=3$: [9.54963;\; -3.10164;\; 1.10588;\; -9.54963]
    } \\
    \hline
    [2; 4; 0] & \makecell{
        $i=1$: [5.42615;\; -6.42016;\; 0.33964;\; -8.93140] \\
        $i=2$: [-8.93140;\; -8.93140;\; 7.94249;\; -1.93755] \\
        $i=3$: [5.44492;\; -6.42955;\; 1.18064;\; -8.93140]
    } \\
    \hline
    [-1; -1; -1] & \makecell{
        $i=1$: [9.33537;\; 9.33537;\; 2.56583;\; -1.66463] \\
        $i=2$: [-6.46393;\; 9.33537;\; -6.38989;\; 9.33537] \\
        $i=3$: [8.81748;\; 9.33537;\; 2.73168;\; -9.33537]
    } \\
    \hline
    [0; 0; 5] & \makecell{
        $i=1$: [9.39358;\; 9.39358;\; -5.01571;\; -6.87332] \\
        $i=2$: [4.58877;\; 9.39358;\; -9.39358;\; -6.30819] \\
        $i=3$: [9.39358;\; -9.39358;\; 0.01605;\; 4.17240]
    } \\
    \hline
    [1; 7; -3] & \makecell{
        $i=1$: [-3.04607;\; 2.89641;\; 9.71846;\; -8.74574] \\
        $i=2$: [-9.71846;\; 0.50025;\; 0.58349;\; -4.78639] \\
        $i=3$: [-9.71846;\; 0.38057;\; -0.34336;\; -2.84356]
    } \\
    \hline
    [1; -4; 0] & \makecell{
        $i=1$: [7.61593;\; 8.86772;\; 0.46901;\; -2.13228] \\
        $i=2$: [3.03558;\; 5.35980;\; -8.86772;\; 4.21875] \\
        $i=3$: [8.51471;\; 8.22830;\; -0.82629;\; -8.86772]
    } \\
    \hline
    \textbf{$x_t$} & \textbf{$[c_i, d_i]$ for the minimum functions} \\
    \hline
    [0; 4; 0] & \makecell{
        $i=1$: [9.68768;\; -8.15707;\; 0.25442;\; -9.68768] \\
        $i=2$: [7.62849;\; -7.31980;\; -0.82692;\; -9.68768] \\
        $i=3$: [9.68768;\; -7.21545;\; 0.91940;\; -9.68768]
    } \\
    \hline
    [2; 4; 0] & \makecell{
        $i=1$: [0.42962;\; -4.40086;\; -0.21866;\; -8.39663] \\
        $i=2$: [-5.12913;\; 0.50961;\; 0.50961;\; -8.39663] \\
        $i=3$: [8.39663;\; 8.39663;\; 1.41919;\; -2.60337]
    } \\
    \hline
    [-1; -1; -1] & \makecell{
        $i=1$: [-1.21793;\; 4.12835;\; 8.92794;\; -3.73406] \\
        $i=2$: [0.44406;\; 4.43967;\; 8.92794;\; -2.07206] \\
        $i=3$: [8.92794;\; 8.92794;\; -0.23395;\; -2.75007]
    } \\
    \hline
    [0; 0; 5] & \makecell{
        $i=1$: [9.42740;\; 5.67547;\; -1.43236;\; -3.24214] \\
        $i=2$: [9.36329;\; 8.58652;\; 0.33968;\; -1.57260] \\
        $i=3$: [9.42740;\; 9.42740;\; 1.15424;\; -1.57260]
    } \\
    \hline
    [1; 7; -3] & \makecell{
        $i=1$: [0.41272;\; -5.86512;\; 0.41272;\; -8.70149] \\
        $i=2$: [0.41272;\; -5.86512;\; 0.41272;\; -8.70149] \\
        $i=3$: [8.70149;\; -3.87139;\; 0.61696;\; -8.70149]
    } \\
    \hline
    [1; -4; 0] & \makecell{
        $i=1$: [3.38486;\; 9.82622;\; 6.33728;\; -1.17378] \\
        $i=2$: [9.82622;\; 0.07198;\; 0.07198;\; -1.17378] \\
        $i=3$: [2.57865;\; 9.82622;\; 5.65450;\; -1.17378]
    } \\
    \hline
\end{tabular}
\end{table}

\begin{table*}[t]
    \centering
    \caption{Results for Example \ref{Ex:Car}: Coefficients (have been rounded down until 2 digits) of polyhedral maximum functions corresponding to each test point }
    \label{tab:Ex-Car}
    \begin{tabular}{|c|c|c|c|}
        \hline
        \textbf{$x_t$} & \textbf{$[c_i, d_i]$ for the maximum functions} & \textbf{\#Cutoff Depth} & \textbf{Computational time (sec)}\\
        \hline
        [-5;-5;-5;-20;-20;-20]  & \makecell{ $i=1: [0.0;0.0;0.29;0.0;0.0;0.68;-9.89]$ \\ $i=2: [ 9.89; -4.20; -4.44; 9.89; 9.89; 9.89; 9.89]$  } & 10  & 2.77\\
        \hline
        [-5;-5;5;-20;-20;-20]  & \makecell{ $i=1: [ 0.0; 0.31; 0.0; 0.0; 0.0; 0.67; -9.88]$ \\ $i=2: [-9.88; 9.88; -3.49; 9.88; 9.88; 7.66; 9.88]$  } & 10 &  2.22\\
        \hline
        [-5; -5; 5; 20; -20; -20] & \makecell{ $i=1: [0.0; 0.0; -0.25; -0.68; 0.0; 0.0; -9.89]$ \\ $i=2: [  9.89; -1.76; -3.31; -9.89; 9.89; 9.89; 9.89]$} & 10 & 10.55 \\
        \hline
        [-5; -5; 5; 20; -20; 20] & \makecell{ $i=1: [0.0; 0.26; 0.0; -0.68; 0.0; 0.0; -9.89]$ \\ $i=2: [9.26; -9.89; 9.89; -2.39; 9.89; 9.89; 9.89]$} & 10 & 13.02\\
        \hline
        [-5; 5; 5; 20; -20; 20] & \makecell{ $i=1: [0.0; -0.13; 0.0; -0.70; 0.0; 0.0; -9.89]$ \\ $i=2: [-9.89; 9.89; -9.44; -9.89; 4.37; -9.89; 9.89 ]$} & 10 & 39.84\\
        \hline
        [-5; 5; 5; -20; 20; 20] & \makecell{ $i=1: [0.0;0.0;-0.30;  0.0;0.0;-0.67;-9.89]$ \\ $i=2: [5.17;9.89;9.32;9.89;-9.89;  9.89;9.89]$} & 12 & 456.16\\
        \hline
        [5; 5; 5; 20; 20; 20] & \makecell{ $i=1: [0.0; -0.3; 0.0; -0.67;0.0; 0.0; -9.89]$ \\ $i=2: [1.00; -4.70; -9.89; -9.89; 9.89; 3.64; 9.89]$} & 14 & 361.05\\
        \hline
         \end{tabular}
\end{table*}

\end{document}